\documentclass[]{elsarticle}

\makeatletter
\def\ps@headings{%
\def\@oddfoot{}%
\def\@evenfoot{}}
\makeatother
\usepackage{epsfig,latexsym,epsf,color}
\usepackage{graphicx,amsmath,amssymb,latexsym,epstopdf}
\usepackage{subfigure}
\usepackage{paralist}
\usepackage{bm}
\usepackage[nomarkers,nolists,tablesfirst]{endfloat}

\usepackage[only,tensf,]{rawfonts}

\newcommand{\tx}[1]{\text{tx}(l)}
\newcommand{\rx}[1]{\text{rx}(l)}

\newcommand{\ie}{{\it i.e.}}
\newcommand{\eg}{{\em e.g.}}

\newcommand{\separator}{
  \begin{center}
    \rule{\columnwidth}{0.3mm}
  \end{center}
}

\newenvironment{separation}
{ \vspace{-0.5cm}
  \separator
  \vspace{-0.3cm}
}
{
  \vspace{-0.4cm}
  \separator
  \vspace{-0.4cm}
}

\newcommand{\bi}{\begin{itemize}}
\newcommand{\ei}{\end{itemize}}
\newcommand{\be}{\begin{enumerate}}
\newcommand{\ee}{\end{enumerate}}

\newcommand{\beq}{\begin{eqnarray*}}
\newcommand{\eeq}{\end{eqnarray*}}
\newcommand{\beqn}{\begin{eqnarray}}
\newcommand{\eeqn}{\end{eqnarray}}

\newcommand{\mc}[1]{\mathcal{#1}}

\newcommand{\set}[1]{\mc{#1}}

\journal{Computer Networks}

\begin{document}

\begin{frontmatter}

\title{Embedding of Virtual Network Requests over Static Wireless Multihop
  Networks}

\author[KAIST]{Donggyu Yun}
\ead{dgyun@lanada.kaist.ac.kr}

\author[KAIST]{Jungseul Ok}
\ead{ockjs@lanada.kaist.ac.kr}

\author[LIG]{Bongjhin Shin}
\ead{bongjin.shin@lignex1.com}

\author[LIG]{Soobum Park}
\ead{sbpark93@lignex1.com}

\author[KAIST]{Yung Yi\corref{cor1}}
\ead{yiyung@kaist.edu}

\cortext[cor1]{Corresponding author, Tel.: +82 42 350 5486}

\address[KAIST]{Department of Electrical Engineering, KAIST,  Daejeon, South Korea }
\address[LIG]{Communication Lab., LIG Nex1, Seongnam-City, Gyeonggi-do, South Korea}

%\author{Donggyu Yun, Jeongseul Ok, and Yung Yi\thanks{D. Yun, J. Ok and
%    Y. Yi are with the Department of Electrical Engineering, KAIST
 %   (Korea Advanced Institute of Science and Technology), South Korea,
  %  Emails: \{dgyun,ockjs\}@lanada.kaist.ac.kr, yiyung@kaist.edu.}  }

% \author{\IEEEauthorblockN{Donggyu Yun, Jeongseul Ok, and Yung Yi}
%   \IEEEauthorblockA{Department of Electrical Engineering\\
%     KAIST (Korea Advanced Institute of Science and Technology), South
%     Korea\\
%     Emails: \{dgyun,ockjs\}@lanada.kaist.ac.kr, yiyung@kaist.edu}
%   }
% make the title area

\begin{abstract}
  Network virtualization is a technology of running multiple
  heterogeneous network architecture on a shared substrate network.
  One of the crucial components in network virtualization is virtual
  network embedding, which provides a way to allocate physical network
  resources (\eg, CPU and link bandwidth) to virtual network
  requests. Despite significant research efforts on virtual network
  embedding in wired and cellular networks, little attention has been
  paid to that in wireless multi-hop networks, which is becoming more
  important due to its rapid growth and the need to share these networks
  among different business sectors and users. In this paper, we
  first study the root causes of new challenges of virtual network
  embedding in wireless multi-hop networks, and propose a new embedding
  algorithm that efficiently uses the resources of the physical
  substrate network. We examine our algorithm's performance through
  extensive simulations under various scenarios. Due to lack of
  competitive algorithms, we compare the proposed algorithm to five
  other algorithms, mainly borrowed from wired embedding or artificially
  made by us, partially with or without the key algorithmic ideas to
  assess their impacts.

% under various scenarios
% by comparing
%   withing
% based on
%   comparison to the algorithms borrowed from wired embedding
% comparing to

%   the performance our

% We evaluate our algorithm to the algorithms (without
%   considering the wireless multi-hop features) through extensive
%   simulations, and demonstrate that about 40\% increase in revenue is
%   achieved. This paper also motivates the research community to pay more
%   attention to embedding algorithm in wireless multi-hop networks.

%   an efficient use of underlying network resources.. The \emph{virtual
%     network(VN) embedding problem} deal with that issue. The objective
%   in VN embedding problem is to find an efficient way to embed each VN
%   elements to the particular nodes and paths in the substrate network.
%   Although researches on VN embedding problem has been mainly carried
%   out in wired network domain, it is also worthy of condsidering this
%   problem in wirelss domain due to the rapid growth of wireless network
%   infrastructure.  However, link interference in the wireless network
%   makes wireless VN embedding problem more challengeable.  In this
%   paper, we propose an efficient wireless VN embedding method for static
%   VN requests.  Our simulation results show that the proposed algorithm
%   enable us to accept much more virtual network requests, compared with
%   a naive method.

\end{abstract}


\end{frontmatter}

\section{Introduction}

% \subsection{Motivation}
Network virtualization is a powerful technology that allows multiple
heterogeneous network architectures over a shared physical network,
shortly called a {\em substrate network} (SN).  Major applications
include testing of new network protocols, such as PlanetLab
\cite{planetlab} and Emulab \cite{emulab}.
%For example, PlanetLab currently consists of
%world-wide hosts over 600 experimental virtualized services.
% For example, PlanetLab is a global research network for
%evaluating and deploying experimental network services and protocols,
%currently consisting of world-wide hosts over 600 experimental
%virtualized services. 
Network virtualization also enables multiple service providers to offer
customized services with different requirements and features, \eg, a
streaming video network with low delay and high bandwidth or a financial
service network with high reliability and security guarantees, over a
common physical substrate network.

% \cite{fea07}
In a virtualized network environment, the SN providers accept and run
multiple virtual networks (VNs), where the substrate nodes serve the
nodes in a virtual network and the physical paths configure the
virtual links. Since multiple VNs are bound to share the common underlying physical
resources, the \emph{VN embedding problem} that finds an efficient
embedding of each VN to substrate resources, is very important in order
to support as many requests as possible and/or minimize the substrate
network resources for embedding. 

% The
% interferences among links, as seriously studied in the area of link
% scheduling [CITE], make it hard to conjecture the amount of available
% resources in the substrate network, when we embed new VN requests, and thus finding an efficient
% embedding becomes significantly challenging. Those challenges include
% addition computational complexity in search for a good embedding,
% requiring an efficient heuristic algorithm to minimize the resource
% usage and maximize the success ratio of embedding new VN requests.

In this paper, we consider a VN embedding problem over wireless
multi-hop networks. We focus on the key difference--inter-link
interference--in wireless multi-hop networks from wired networks, whose
challenges are summarized next.  A typical embedding algorithm consists
of the following procedures:
% (see
% Fig.~\ref{fig:flowchart}): 
{\em (a)} search plausible candidate
embeddings for a given VN request, {\em (b)} assess their qualities in
terms of a target objective, \eg, minimization of substrate resource
usage or revenue maximization, and {\em (c)} select the best candidate
among the feasible ones\footnote{We say that an
  embedding is feasible when the physical substrate network has enough
  resource to support the embedding.}.
% This repetition is
%often limited by some numbers to generate an embedding solution for a
%given temporal or computational budget in practice.
% \begin{figure}[h!]
%   \centering
%   \includegraphics*[width= 6cm]{flowchart}
%   \caption{Embedding process for a VN request}
%   \label{fig:flowchart}
% \end{figure}
From the above procedures, we see that two primitives are necessary in embedding:
{\em (i)} checking the feasibility of a candidate embedding and {\em (ii)}
quantifying the embedding's quality. Note that these two primitives are
easy to support in wired networks.  Feasibility can be checked by
comparing the physical layer's remaining resource with the resource
required by the existing and new embeddings. The quality of the embedding is
typically quantified by computing the total amount of resource occupied
by the embedding. There, embedding challenge is due to
computational intractability just coming from the large search
space of candidate embeddings.

%The challenge of embedding is due to computational intractability coming from the large search space of candidate embeddings
%

However, in wireless embedding, both primitives are harder to
check, and even need to be defined appropriately. This is because {\em
  the resource allocated over a link makes an indirect impact on
  the actual remaining resource over other neighboring links due to
  inter-link interference, and both primitives are coupled with the
  underlying MAC.} A MAC protocol (\eg, 802.11), has its unique capability of
supporting the assigned rates over each link, meaning that the feasibility
check will lead to different results, depending on the underlying MAC.
Regarding the quality comparison metric, as an example, consider two
embeddings $E_1$ and $E_2,$ where the aggregate amount of resource
required by $E_1$ exceeds that by $E_2.$ However, $E_1$ can be
preferable if the embedded nodes and links in $E_1$ are in less
interfering regions, because it is likely that more future requests will
be accepted. 

Our approach and main contributions towards efficient embedding in
wireless multi-hop networks are summarized as follows:

\smallskip
\begin{compactenum}[1)]

\item {\em Feasibility check.}  We propose two solutions: {\em (i)}
  sufficiency-based approach and {\em (ii)} simulation-based approach
  with smart embedding. First, in the sufficiency-based approach, we use
  a graph-theoretical sufficient condition based on weighted graph
  coloring, which, if met, feasibility is guaranteed. Second, in the
  simulation based-approach, we limit the space of candidate
  embeddings, so that the conflict graph\footnote{It is the graph that
    represents interference relationships among wireless links. Refer to
    \ref{sec:feasibility_checking} for a more detailed explanation.} of the embedded substrate network\footnote{It is a
    subgraph of the substrate network consisting only of nodes and links
    which serve some virtual network requests.} always satisfies a specific
  pattern, called PBG (Polynomially Bounded Growing) graph. To
  check the feasibility of a candidate embedding, we simulate the substrate
  network, and the PBG property enables us to check the feasibility polynomially and
  performs arbitrarily close to the optimal one.  

% {\bf DG: In this paper, we introduce two ways of confirming the feasibility of
% certain embedding within a proper time. The first one is adopting a
% sufficient condition for feasible embedding. The sufficient condition is
% a suitably modified one for the graph coloring problem.  The other way
% is running simulation the substrate network with respect to an embedding
% and observing whether the substrate network is supportable. Especially,
% it is revealed that running approximated MWIS (Maximum Weight
% Independent Set) scheduling yields an approximated solution for the
% feasibility checking problem.  Since finding an approximated MWIS is
% even hard in general case, we use some technique to make it can be
% practically solved.}

\item {\em Quality comparison metric.}  We choose a simple quality
  metric (for a candidate embedding) that is designed to decrease when less overall
  loads are imposed on the SN by the embedding.  While VN nodes are
  one-to-one mapped to the SN nodes, a VN link can be mapped to a path
  in the SN and the metric is designed so that the amount of
  link-interference of in the path is minimized. 
% We use a simple
%  combination of the normalized loads of the link capacities and the
%  degrees of the links.
% This metric is parameterized so that the metric
 % captures the designer's choice and emphasis on selecting good
  %embeddings.

% To
%   quantify it, we use $|| \bm{\lambda} \cdot \bm{D}||_p,$ where
%   $\lambda_l$ is the normalized loads, and $$

% Since nodes in the same
%   VN are basically mapped one-to-one onto SN nodes, There are no big
%   difference among overall loads on SN nodes with respect to candidate
%   embeddings. However, a VN link can be mapped to a SN path not only a
%   link and traffic loads on a substrate link restrict the operation of
%   other links that interfere with. Hence it is important how much loads
%   on which links will be increased.  From this reason, we evaluate
%   embedding's quality based on the weighted loads on SN links, which is
%   reflected interference relationship among SN links.

\item {\em Efficient candidate searching.} The key to an efficient
  embedding algorithm lies in how to smartly search a limited set of
  ``good'' candidate embeddings. We repeatedly test a candidate
  embedding that is chosen by merging the node and link mapping process
  for a limited number of times. The joint link and node
  mapping process simultaneously considers the amount of available node
  resources in the node mapping and the degree of newly generated
  interference to the network in the link mapping. This selection of
  nodes and links are coupled with the comparison metric.

%  in a
%   coupling with the defined:

% We use the defined metric to develop a
%   final embedding algorithm. 
%   a) Select 

% Good embedding algorithms are completed
%   by testing good embedding candidates. The key of our approach is to
%   combine node and link embedding process: Select the SN node hosting a
%   VN node by considering two end points of substrate paths that will
%   serve the virtual links connected to the VN node. We find the optimal
%   path on the metric of embedding quality for given two end points, and
%   then conversely we can calculate which end points obtain the best path
%   among such optimal paths, it essentially decides the SN node for a VN
%   node.

 % \item {\em Performance evaluation.} 

 % \vspace{3cm}

\end{compactenum}
\smallskip

Potential applications of VN embedding over wireless multi-hop networks
are as follows: With increasing number of mobile users, accelerated by proliferation of
smart phones, wireless access technologies are becoming diverse,
widespread, and broadband. Of many types of access networks, wireless
multi-hop networks, are expected to be used as an inexpensive way to
provide last-mile Internet access.  In fact, several cities are
currently deploying municipal wiress mesh networks \cite{hud10}.  Virtualization can
serve plenty of uses over multi-hop networks too.  In the mobile network
market, a growing number of Mobile Virtual Network Operators (MVNO),
that reaches over 430 worldwide in 2010 \cite{mvno}.  MVNOs do not own
the wireless network infrastructure, and lease part of the physical
infrastructure from Mobile Network Operators (MNO) to provide customized
mobile services, in which case MNOs having single/multi-hop wireless
physical networks may need to virtualize their networks for the business
with MVNOs. In addition, similar to PlanetLab, Orbit \cite{orbit} which
is a wireless network testbed consisting of 20$\times$20 nodes can be an
another example of virtualization over the wireless multi-hop network.
Besides above applications, there will be other futuristic scenarios in
which multi-hop virtualization is utilizaed, when wireless multi-hop
networks come into wide use.

% \smallskip
% \noindent{\bf \em Limitation.} 
We consider only inter-link interference modeled based on a
graph-theoretic relationship, and do not consider wireless links'
time-varying characteristics due to e.g., SN nodes' mobility, \ie, the
capacities of links are assumed to be fixed. Although this does not
reflect the practice perfectly, our work can be an important step
towards efficient embedding over wireless multi-hop networks, since
handling inter-link interference is one of the major obstacle there,
such as the approach in wireless link scheduling research (see e.g.,
\cite{yichi08b} for a survey). We expect that our work is connected to
research on more practical algorithms reflecting the full wireless
features in the future.

\section{Related Work}
\label{sec:related}

Recently, there has been research interest regarding virtual network
embedding over wired networks, \eg,
\cite{fan06,lu06,zhu06,yu08,mos09,lis09} and/or embedding in single-hop
cellular networks\cite{kok10}, where the embedding problem turns out to
require computationally intractable complexity for optimality, and thus
various heuristics have been proposed.  More challenging issues in
multi-hop networks are involved mainly due to the complex interference
among links and its severe coupling with network topology. These
challenges require a new design of embedding algorithms.

Recently, NVS in \cite{kok10} is proposed for virtualization on wireless
single-hop networks.  NVS provides an effective wireless resource
allocation over cellular networks, by separating slice scheduling and
flow scheduling.  However, the VN embedding problem essentially asks the
mapping correspondence between a VN node/link and SN node/path, beyond
the scope of resource allocation.

Related work on embedding in wired networks mainly focuses on addressing
computational challenges by restricting the problem space in different
dimensions or proposing heuristic algorithms \cite{fan06,lu06, zhu06}.
For example, only bandwidth requirements are considered in
\cite{fan06,lu06} or all VN requests are assumed to be given in advance
\cite{lu06,zhu06}.  The authors in \cite{fan06,lu06,zhu06} also
considered the substrate network with infinite capacity, accepting all
incoming VN requests.  On embedding problems over SN with limited
resources, multicommodity flow detection based algorithms are proposed
in \cite{yu08,mos09}, where node embedding methods also consider their
relation to the link embedding stage like our work. However, all
algorithms in \cite{lu06,zhu06,yu08,mos09} separate node and link
embedding process, \ie, all VN nodes are embedded before embedding the
VN links.  A single-stage wired embedding algorithm is also proposed in
\cite{lis09} which tries to find a subgraph isomorphism of the VN via a
\emph{backtracking} method.  There, the algorithm imposes a limit on the
length of substrate paths that will embed virtual links and checks
feasibility whenever a new VN link is embedded: if it is infeasible, it
is backtracked to the last feasible embedding.

We extend the domain of the embedding problem to wireless multi-hop
networks, keeping considerations for more general cases such as online
requests and SN with limited capacity.

% there are several work for new
%platform designs \cite{smi07,mah08,sing08}. A TDM-based (Time Division
%Multiplexing) wireless virtualization is introduced in \cite{smi07},
%whereas the authors in \cite{sing08} take an FDM-based (Frequency
%Division Multiplexing) scheme. Recently,
%NVS is proposed in \cite{kok10}, which provides an effective wireless
%resource allocation over cellular networks, by separating slice
%scheduling and flow scheduling.

%%% Local Variables: 
%%% mode: latex
%%% TeX-master: "main"
%%% End: 

%\input{problem.tex}

\section{Wireless VN Embedding Problem} \label{sec:problem}

\subsection{Model}

% \noindent{\bf \em Substrate network}
% \smallskip
\subsubsection{Substrate network}
%\smallskip

\noindent{\bf Network model.} 
Denote the wireless substrate network (SN) by an undirected graph
$G^{S}=(N^{S},L^{S},A^{S}_N,A^{S}_L,I)$ \footnote{Throughout this paper,
  we use the superscripts $S$ and $V$ to refer to the notations related
  to substrate and virtual networks, respectively. }, where $N^{S}$ and
$L^{S}$ are the sets of nodes and links, respectively.  The $A^{S}_N$
refers to the set of node resources, \eg, CPU resource or hard-disk.
We assume that only CPU resource is considered in this paper, which can
be readily extended to other resources. The $A^{S}_L$ is the set of link
resources. We restrict our
attention to the case of providing the long-term average capacity to the
link resource requirement in a VN request. Thus, we model the link
capacity to be fixed, which is a time-averaged value over time-varying
link channels. This seems to be reasonable since the time-scale of
embedding arrival and departure is much slower than that of channel
variations. Let $\text{CPU}^{S}(n)$ be the amount of CPU resource of node $n \in
N^{S},$ and $\text{CAP}^{S}(l)$ be the capacity of wireless link $l \in
L^{S}.$ We also denote by $P^{S}$ the set of all paths in $G^S.$

\smallskip
\noindent{\bf Interference model.} The matrix $I$ is the $|L^{S}| \times |L^{S}|$ matrix which
represents the interference relationships for wireless links, where
$I_{ij}=1$ if links $i$ and $j$ interfere with each other, and 0
otherwise.  Denote by $d_l$ the number of interfering links with $l$
in the SN. The interference matrix depends on the physical
layer techniques as well as the employed MAC. In literature on modeling
wireless networks, a hop-based interference model is popularly used,
\eg, one-hop for FH-CDMA and two-hop for 802.11-like systems. However, our
description can be readily extended to the general cases.

\begin{table}[]
 \caption{Summary of Major Notations}
  \begin{centering}
    \begin{tabular}{|c|l|}
      \hline
     \bf{ Notation} & \bf{Description}\tabularnewline
      \hline
      \hline
      $G^{S}$ & substrate network\tabularnewline
      \hline
      $N^{S},L^{S}$ & set of substrate nodes and links \tabularnewline
      \hline
      $\text{CPU}^{S}(n)$ & CPU resources of SN node $n \in N^{S}$\tabularnewline
      \hline
      $\text{CAP}^{S}(l)$ & capacity of SN link $l \in L^{S}$\tabularnewline
       \hline
      $d_l$ & number of interfering links with SN link $l \in L^{S}$\tabularnewline
      \hline
      $G^{V}$ & virtual network\tabularnewline
      \hline
      $N^{V},L^{V}$ & set of virtual nodes and links \tabularnewline
      \hline
      $\text{CPU}^{V}(n)$ & CPU requirement of VN node $n \in N^{V}$\tabularnewline
      \hline
      $\text{BW}^{V}(l)$ & bandwidth requirement of VN link $l \in L^{V}$\tabularnewline
    \hline
\end{tabular}\label{modelnotations}
    \par\end{centering}
  \vspace{0.1cm}
\end{table}

\smallskip
\noindent{\em MAC model.} 
We assume a MAC with {\em $\epsilon$-throughput-optimality} for some
$\epsilon >0.$ Roughly, a MAC is said to be {\em throughput-optimal} if
it can stabilize any arrival rate vector over the SN links whenever
possible (see the seminal paper \cite{tas92} for a formal definition).
$\epsilon$-throughput-optimality means that a MAC can support only a
$\epsilon$-reduced version of the arrival rates supported by a
throughput-optimal MAC. %There exists a large list of researches that
%propose distributed MACs with such performance guarantee.
 We can regard $\epsilon$ as a reduction factor that is due to any kind of
implementation overhead, \eg, message passing. We note that recently
there also exists a research on so-called optimal CSMA, \eg,
\cite{yicsma09,SRS09,qcsma} that is close to throughput-optimal by
simply and locally controlling CSMA parameters. % Note that it does not
% mean that our embedding framework completely excludes the use of
% practical MACs such as 802.11, which we comment later in this
% paper.

% We mainly focus our research on futuristic
% scenarios, such that the underlying MAC is equipped with a performance
% guarantee. This has been extensively studied in literature.  A short
% summary is : Since the seminal paper by Tassiulas and Ephreemides, where
% a throughput-optimal MAC, referred to as Max-Weight scheduling, is
% presented, the distributed MAC protocols with full or partial
% performance guarantee have been proposed. In particular, recently, a
% simple adaptive CSMA can achieve the performance close to that by
% Max-Weight, and the research showing that it can be implemented by a
% simple change of the current 802.11 MAC has been presented [CITE] {\bf
%  DG : Which paper should I refer?}. In
% this trend of bridging the gap between theory and practice, this paper
% targets the embedding study into when such new MACs become mature and at
% least its design philosophy is implemented in the future
% standards. Adopting this ``advanced" MAC also provides convenience of
% understanding and developing embedding algorithms more
% fundamentally. {\bf DG : We need to modify the following sentence.} 

% We consider \emph{online} virtual network requests scenario. It
% means that each VN request arbitrarily arrive and stay in the
% network over time. 

% \smallskip
% \noindent{\bf \em Virtual network}
%\smallskip
\subsubsection{Virtual network}

A virtual network (VN) request is denoted by an undirected graph
$G^{V}=(N^{V},L^{V},C^{V}_N,C^{V}_L,D^{V})$, where $N^{V}$ and $L^{V}$
are the set of nodes and links of the virtual network, respectively. 
$C^{V}_N$ and $C^{V}_L$ denote the set of node and link requirements. As
an example, let $\text{CPU}^{V}(n)$ and $\text{BW}^{V}(l)$ be the CPU
and bandwidth requirements for the virtual node $n \in N^V$ and the
virtual link $l \in L^V$.
 These requirements can be also interpreted as \emph{constraints} for
embedding \ie, to accept a VN request, the amount of allocated substrate
resources for each virtual node and link should be more than
required. $D^{V}$ implies that the VN request should be served for
the duration of $D^{V}.$ With a little abuse of notation, we also denote by $L(n)$ the set of the links connected
to the node $n$ in both $G^V$ and $G^S$.
% depending on whether $n \in N^V$ or $N^S.$ 

\subsection{Problem Formulation: Virtual Network
  Embedding} \label{sec:VNembedding}

We define an embedding $E$ from $G^{V}$ to a subset of $G^{S}$ as a mapping
\begin{displaymath}
E : G^{V}  \to (N^{\star},P^{\star},R^{N},R^{L})
\end{displaymath}
where $N^{\star}$$\subset$$N^{S}$, $P^{\star}$$\subset$$P^{S}$ and
$R^{N},R^{L}$ are the node/link resources allocated to the $G^{V}$
by embedding $E$.

We consider an \emph{online} virtual network requests scenario, where a
sequence of VN requests arbitrarily arrive and stay in the network over
time. We consider a time-slotted system, indexed by $t=0,1,\ldots.$ 

For an embedded VN request $G^V$, the SN provider earns
revenue $R(G^V)$ which is proportional to the total amount of the
requested resources in the VN request, \ie, 
\begin{equation} \label{eq:revenue} R(G^{V}) = \sum_{n\in
    N^{V}}\text{CPU}^{V}(n) + \alpha\sum_{l \in L^{V}}\text{BW}^{V}(l),
\end{equation}
where the constant $\alpha$ reflects the relative importance in node and
link resources, chosen by a virtual network operator. This weighted
method is one of the typical ways to handle multi-objective optimization
problem, like in the previous works \cite{yu08,mos09}.

Then, the SN provider's revenue $R(t)$ at time slot t becomes
\begin{equation} \label{eq:revenue_t}
R(t)=\sum_{g \in \set{G}^{V}(t)}R(g),
\end{equation}
where $\set{G}^V(t)$ be the set of VN requests served at time slot t.
Since VN requests arrive and depart arbitrarily over time, we set a goal
of our embedding algorithm to maximize the time-averaged revenue, given by:
\begin{equation} \label{eq:av_revenue}
  \max \lim_{T \to \infty} \frac{\int_{0}^{T} R(t)}{T}.
\end{equation}

% Optimally solving (\ref{eq:av_revenue}) turns out to be a complex dynamic
% programming, which is hard to compute, due to high complexity and the
% need of knowing future events or stochastic features of VN arrivals, and
% thus we need a heuristic solution.

%%% Local Variables: 
%%% mode: latex
%%% TeX-master: "main"
%%% End: 

%\input{challenge.tex}

\section{Challenges in Wireless Embedding} 
\label{sec:challenge}

\subsection{Handling Online VN Requests}
\label{sec:Online VN embedding}

The VN request arrivals and departures are not known in advance, and may
be quite random. In this unpredictable situation, the on-line embedding
algorithm to achieve the objective (\ref{eq:av_revenue}) requires
the statistical properties of VN requests and the large search space,
where a mathematical tool such as dynamic programming will be used.
Clearly, the issue of handling online, unpredictable VN requests
also exists in the embedding over wired substrate networks. 

% How and which VN requests have been embedded so far determines the
% acceptable future VN requests.  Conversely, if the future VN requests
% are known in advance, a complex dynamic programming can inform us which
% VN requests should be accepted in current step to achieve the goal in
% (\ref{eq:average_revenue}).  However, the dynamic programming method is
% practically inapplicable due to high complexity.  Furthermore, it can be
% established only for known future events, but it is even hard to expect
% upcoming VN requests.  Instead, we can consider a heuristric approach
% for achieving the goal.  It would be a discerning admission control
% mechanism for deciding which VN requests will be served.  For each
% selected VN request, we need an efficient embedding method of a VN
% request.
 
\subsection{Hardness of Embedding Quality Comparison}
\label{sec:embedding_comparison} 

After searching a multiple of candidate embeddings for a given VN, the next step
is to quantify how good the embeddings are. Comparing quality among 
candidate embeddings has the following requirements:

\medskip
\begin{compactenum}[\em R1.]
\item {\bf \em Simplicity.} Short computation time for comparison is
  necessary due to possibility of existence of a large number of
  candidate embeddings. 

\item {\bf \em Efficiency.} The devised metric should appropriately reflect
  the changes in the available substrate resources for future requests. 
\end{compactenum}
\medskip

In wired substrate networks, {\em R1} and {\em R2} are easy to meet,
since the amount of available substrate resource is just the original
amount of the resource subtracted by the resource assigned in embedding
previous requests. However, in wireless multi-hop substrate networks, as
discussed earlier, measuring the amount of the available resource is
ambiguous.  We can interpret the available (or remaining ) bandwidth
resource of a wireless link as its capacity multiplied by the time
portion that it can be activated additionally. Therefore, the amount of
available resource over each link is determined by the employed link
scheduling.  However, wireless networks operate a dynamic scheduling
algorithm which decides a set of activated links over time. This makes
exact calculation of the amount of available resources practically
impossible.

\subsection{Feasibility Checking Problem}
\label{sec:feasibility_checking} 

\begin{figure}[]
  \centering
  \subfigure[VN request]{%
    \label{fig:feas_ex(a)}
    \includegraphics*[width=0.18\columnwidth]{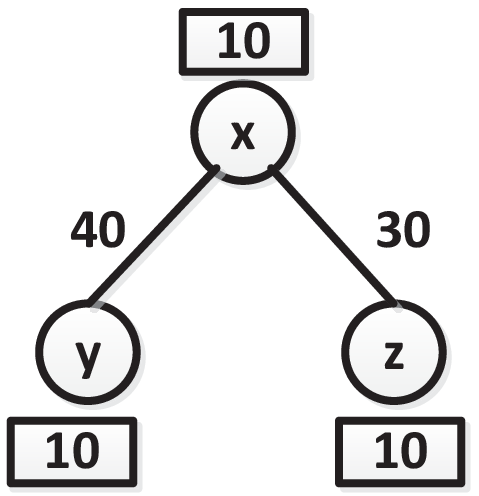}} \hspace{0.2cm}
  \subfigure[SN]{%
    \label{fig:feas_ex(b)}
    \includegraphics*[width=0.33\columnwidth]{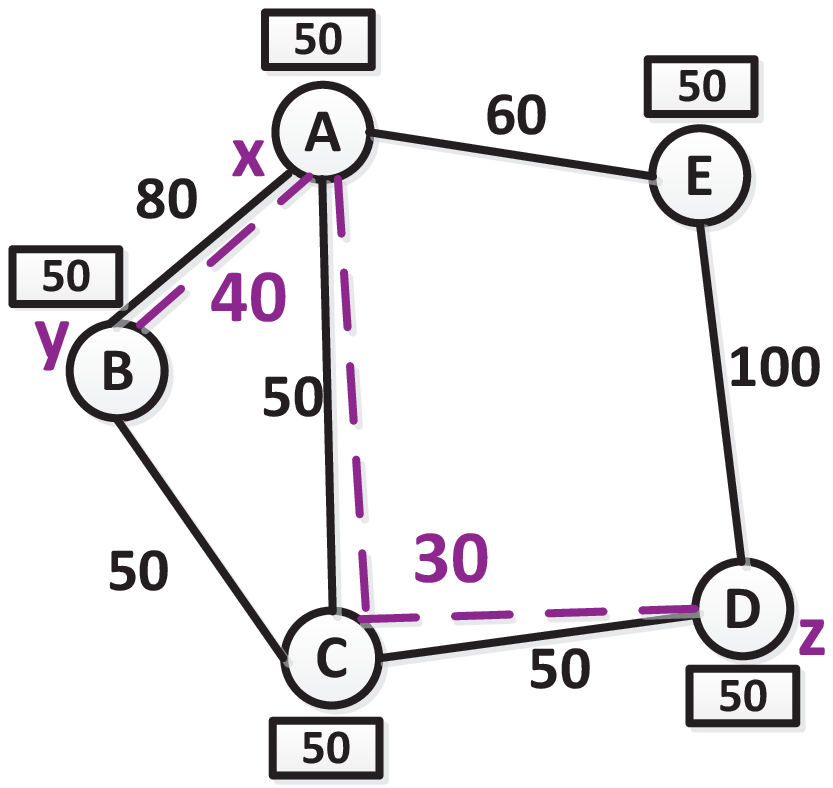}}
  \caption{(a) A VN  request (b) An embedding in SN, The SN is either
    wired or wireless.  If it is wireless, the interference is one-hop based, \eg, links $(A,C)$
    and $(C,D)$ interfere. The numbers in the box and on the
    links represent the amount of CPU resources and link bandwidths. }
  \label{fig:feas_ex}
\end{figure}

% \smallskip
\noindent{\bf Wired vs. Wireless.}
We consider an example in Fig.~\ref{fig:feas_ex} to explain the unique
challenges in the feasibility check of wireless embedding.  Note that
feasibility check is an important primitive in embedding to check if
there is enough resource for a candiate embedding to be supported in the
SN. The SN in Fig.~\ref{fig:feas_ex(b)} can be interpreted either as a
wired SN or a wireless SN.  We assume that no prior VN requests are
served in the SN, and the interference in the wireless SN is one-hop
based, \ie, any two links with one distance interfere. Consider a
candidate embedding in both of wired and wireless SNs in which
\begin{eqnarray*}
  \text{\em Node mapping}: && x \rightarrow A, \ y \rightarrow B, \ z
  \rightarrow D \cr
  \text{\em Link mapping}: && (x,y)
\rightarrow (A,B), \ (x,z) \rightarrow (A,C,D),
\end{eqnarray*}
Recall that $(A,C,D)$ is a path in the SN. In the wired SN, feasibility
check can be easily done by individually checking the feasibility of
node and link resources. Although the mapping is feasible in wired SN,
the same embedding is {\em infeasible} in the wireless SN, because there
is no way of serving either $(A,C)$ and $(C,D)$ at one instant and
providing the (long-term) bandwidth of 30 at each link.  This is because
link $(A,C)$ and link $(C,D)$ should be acted as the virtual link
$(x,z)$ during 0.6(=$\frac{30}{50}$) portion of the time, but they
cannot be activated simultaneously due to interference.  This shows the
difficulty of checking feasibility, where individual resource check at
nodes and links is not enough, and a more complex checking procedure
should be considered.

% a wired substrate network
% and a wireless substrate network.  In the VN request, the numbers in the
% boxes represent the requested amounts of CPU resources and the numbers
% beside links represent the requested amounts of bandwidth resources.
% The numbers in the substrate networks represent the amounts of available
% CPU resources and bandwidth resources.  In case of the embedding in
% Fig.\ref{fig:feasibility}(b) all substrate nodes and links have enough
% resources to serve their corresponding virtual nodes and
% links. Therefore the embedding is feasible over the wired substrate
% network.

% There is also the same embedding example in wireless substrate network
% in Fig. \ref{fig:feasibility} (c). The interference relationship is
% represented using red dotted lines.

% Let's look at the green embedding in Fig. \ref{fig:feasibility} (c).
% Virtual link $(x,z)$ should be operated as if it has 30 bandwidth
% resource units. In order to satisfy the requirement, each of link
% $(A,C)$ and link $(C,D)$ should be acted as the virtual link $(x,z)$
% during 0.6(=$\frac{30}{50}$) portion of the time.  However it is
% impossible scenario, becausee only one of two links can be activated at
% any time.  On the other hand, we can see that another embedding
% (displayed in violet) is a feasible one.

\smallskip
\noindent{\bf Formalism.}
Fundamental root causes of feasibility check in wireless embedding can
be understood more clearly by introducing a notion of {\em conflict
  graph}. A conflict graph graphically captures the interference relation
between any pair of links by transforming the original graph for a
given interference matrix such as: a link becomes a node and two links are
connected if they interfere. For a candidate embedding, consider the
{\em potential normalized load} of a substrate link $l$, $\lambda_l =
\text{Req}(l)/\text{CAP}^S(l)$, where $\text{Req}(l)$ is the required
aggregate bandwidth of VN requests already being served and a new
candidate embedding. Then, the candidate embedding becomes feasible if
there exists a scheduling scheme which provides
the long-term average rates of $\lambda_l.$ This has been traditionally
studied in the context of weighted graph coloring for the conflict graph, known to be an
NP-hard problem.  Another way of understanding this problem from the
control and networking theoretic perspective is whether the
system can be stabilized or not by some scheduling algorithm,
assuming stochastic packet arrivals with mean $(\lambda_l)$ over each
link \cite{tas92}.

\subsection{Searching candidate embeddings} 
Section~\ref{sec:Online VN embedding} discussed the challenge of
unpredictable on-line requests. Sections~\ref{sec:embedding_comparison}
and \ref{sec:feasibility_checking} deal with the issue of comparing
embeddings' quality and checking feasibility.
The final embedding algorithm remaining is to determine which of the candidate embeddings to
try. This is because the entire searching of all candidates embeddings
is computationally impossible. Obviously, this problem also appears in
the wired embedding, yet it is coupled with feasibility check as well as
quality comparison metric. Thus, based on the appropriate comparison
metric definition and the feasibility checking mechanism, the question
which embedding to try first and which embedding is finally selected
will be decided accordingly.

% \ref{}

% The challenge in Sections~\ref{sec:feasibility_checking} and
% ~\ref{sec:embedding_comparison} are about given embedding candidates.

% Since embedding candidates being considered directly lead to the
% performance of algorithm, making good embedding candidates is crucial
% component of embedding process.  An embedding instant that is highly
% praised by the embedding comparison metric (in Section
% ~\ref{sec:embedding_comparison}) deserve to become one of embedding
% candidates.  However, it would be hard to find the optimal embedding
% candidate with respect to the criteria. This is because VN embedding
% problem is basically finding the optiaml embedding for a certain
% objective and many previouse VN embedding problem with different
% objectives are known as NP-hard.  If so, we need a heuristic method in
% which features of the comparison metric are fully considered.

%%% Local Variables: 
%%% mode: latex
%%% TeX-master: "main"
%%% End: 

%\input{solution.tex}

\section{Embedding Algorithm}
\label{sec:solution}

\subsection{Algorithm Overview}

\begin{figure}[]
  \centering
  \includegraphics*[width=0.85\columnwidth]{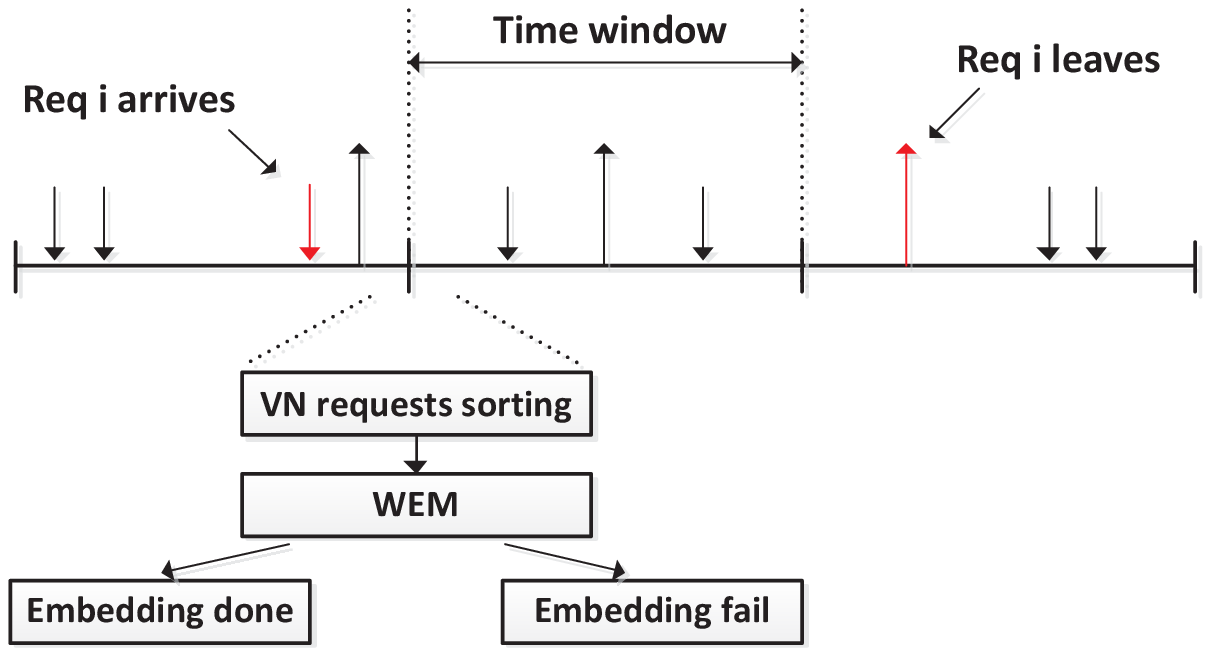}
  \caption{Embedding framework for dynamic VN requests}
  \label{fig:framework}
\end{figure}

The framework of our algorithm is described in Fig.~\ref{fig:framework}
and {\bf WEM} (Wireless EMbedding). First, we explain how dynamic
requests are processed, and then elaborate on the embedding algorithm
for a VN request in the subsequent subsections, which is our major focus
of this paper.

\smallskip
\noindent{\bf On-line requests.}  
Serving incoming VN requests as fast as possible may be one criterion of
handling on-line requests. However, a more crucial capability required
by SN is to prevent the VN request from being blocked due to inefficient
SN resource management. To that end, we divide the time into a sequence
of windows whose duration determines how frequently the embedding
process is run. This is dependent on the SN provider's operation
decision, \eg, an hour or a day.  Over one time window, we collect a
group of arriving requests and process one-by-one.
% Since the amount of SN resources is limited,
Since it may be impossible to accomodate all the incoming requests
within a certain time window, we need to smartly decide the requests
that will be served.  We take an approach that the SN provider may
prioritize the VN requests to maximize the earned revenues by serving
the requests in the order of revenues. This approach is similarly taken
in wired embedding \cite{yu08}. 

\smallskip
\noindent{\bf Algorithm description for a VN request.}  
Once the sequence of VN requests to be processed in a time window is
determined, we try to maximize the efficiency of using SN resources by
employing a smart embedding algorithm for a VN request.  We now describe
our algorithm, called {\bf WEM}, the embedding algorithm for a VN
request as follows:  

\medskip
\begin{separation}
  {\bf WEM: Wireless EMbedding algorithm for a VN request}
  \vspace{-0.4cm} \separator
  \begin{compactenum}[\bf Step 1.]
    \vspace{-0.1cm}
 
  \item Decide the embedding order of VN nodes.

  \item Select the $K$ SN ``root'' nodes, each of which maps the first
    VN node (chosen by {\bf Step 1}) in the $K$ candidate embedding trials.

  \item Choose $K$ candidate embeddings, where each candidate embedding
    process starts from each root node (chosen in {\bf Step 2}) and
    sequentially embeds other remaining VN nodes and links according to
    the order in {\bf Step 1}.

  \item Evaluate $K$ candidate embeddings based on a
    comparison metric, and take the highest-quality embedding candidate
    that passes feasibility check.
    
  \end{compactenum}
\end{separation}

\bigskip
\noindent{\bf Key ideas.} 
Our algorithm is designed to ensure to tackle the challenges in
Section~\ref{sec:challenge}. Scarce resources in wireless networks
should be efficiently utilized, whereas the impact of wireless
interference in many algorithmic components are appropriately handled
while striking a balance between efficiency and running time. In our
algorithm, we first choose $K$ embeddings ({\bf Step 3}), where $K$ is
the search number, and in each candidate embedding, we choose the ``{\em
  root}'' SN node ({\bf Step 2}) that maps to the first VN node (which is
chosen by a selection rule, {\bf Step 1}), Afterwards, VN links and
nodes are mapped simultaneously. This is because interference is coupled
with both nodes and links, \eg, a node with many incident links is
likely to be in the region with severe interference. Finally, we choose
the one that has the best quality and passes the feasibility check ({\bf
  Step 4}). We now provide the details in each step in conjunction with
the design rationales. To help the readers to understand, we will use
the example in Fig.~\ref{fig:ex_input} in all explanations.

\subsection{VN Node Sequence and SN Node Selection}

\noindent{\bf VN node sequence.}
In {\bf Step 1,} we first determine the mapping sequence of VN nodes in
any embedding candidate. To that end, we define the notion of {\em
  extended required node resource}: for a VN node $n,$ $
\text{CPU}^{V}(n) + \alpha \sum_{l\in L(n)}\text{BW}^{V}(l).$
% \begin{eqnarray*}
% D(n^{V}) & = & \text{CPU}^{V}(n^{V}) + \alpha \sum_{l\in L(n^V)}\text{BW}^{V}(l),
% \end{eqnarray*}
This notion captures
the CPU resource plus the aggregate bandwidth resource required by the
node's incident links. The node with highest extended required resource is
firstly embedded due to higher embedding difficulty. We embed the
rest of the nodes in ascending order of hop distance from the first node. 

\smallskip
\noindent{\bf SN node selection.} In {\bf Step 2,} we choose a starting root node in SN for each candidate
embedding. Since we try $K$ embeddings, $K$ starting root nodes should
be selected. We use a similar metric to that in {\bf Step 1}. We sort all
the SN nodes in the decreasing order of the following {\em extended remaining
  node resource}: for an SN node $n$, ${\text{CPU}}^S_{\text{Rem}}(n) +
\alpha \sum_{l\in L(n)}\text{BW}^{S}_{\text{Rem}}(l),$ where
$\text{CPU}^S_{\text{Rem}}(n)$ is the {\em remaining} CPU resource
available for new VN requests (similarly,
$\text{BW}^{S}_{\text{Rem}}(l)$). The extended remaining node
resource quantifies the amount of remaining node resource considering
the bandwidth resource of connected links to the corresponding node.  It
means that we prefer to try the root node which has enough resource.

% $K$ SN nodes each of which will act as a root
% node in each K candidate embeddings. We design this part to search the
% various regions in the SN that leads to the finding of a final embedding
% that is efficient. In WEM, we sort the SN nodes in the descending order
% of {\em } $Q(n^S)$, defined by:
% \begin{eqnarray*}
% Q(n^{S}) & = & \text{CPU}^{R}(n^{S}) + \alpha \sum_{l^S\in L(n^S)}\text{BW}^{R}(l^S),
% \end{eqnarray*}
% where $\text{CPU}^{R}(n^{S})$ is the amount of residual CPU resources of
% $n^S$ and $\text{BW}^{R}(l^{S})$ is the amount of residual bandwidth resources of
%  $l^S$, which is $\text{CAP}^{S}(l^{S})\cdot(1-\lambda_{l^{S}})$. {\bf
%    YY: I don't understand this. Explain the rationale and example}. 

\begin{figure}[]
  \centering
  \subfigure[VN request]{%
  \label{fig:ex_input(a)}
  \includegraphics*[width=0.2\columnwidth]{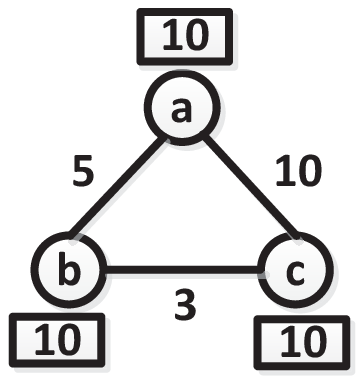}}
  \subfigure[Wireless SN]{%
  \label{fig:ex_input(b)}
  \includegraphics*[width=0.33\columnwidth]{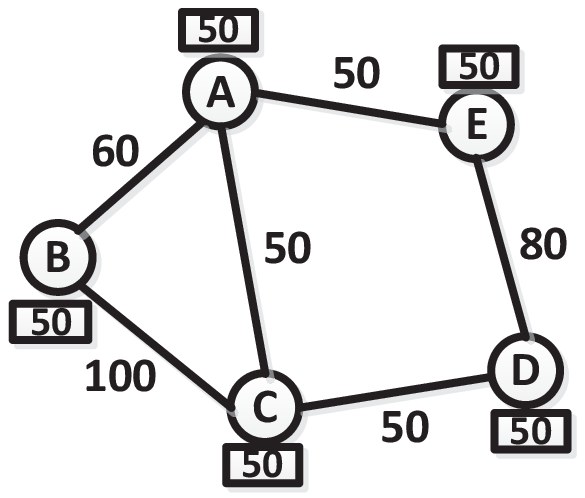}}
  \subfigure[Wireless SN with influence weight]{%
  \label{fig:ex_input(c)}
  \includegraphics*[width=0.36\columnwidth]{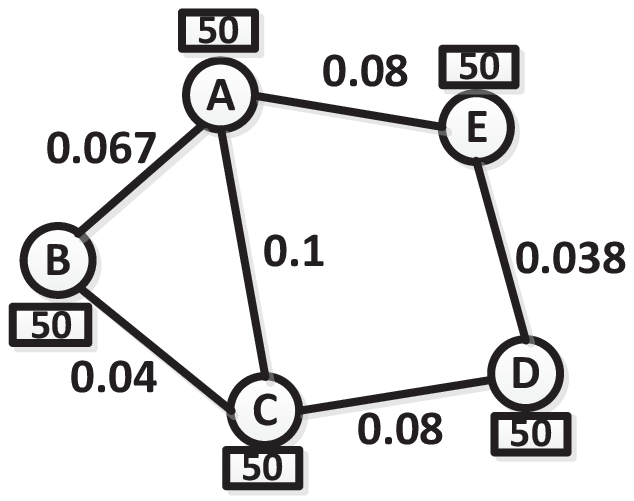}}\\
 \caption{An example of VN and SN to explain WEM. No
   existing VNs in service and one-hop interference
   model are assumed in the SN.}
\label{fig:ex_input}
\end{figure}

\medskip
\noindent{\bf Example.} In Fig.~\ref{fig:ex_input(a)}, since node $a$ has the highest extended
required resources, the embedding sequence is $a, c, b$. Both 
$b$ and $c$ are one hop apart from $a$, here, we give a higher priority to
$c$ because its extended required resources is higher than $b$'s. 
It can be easily shown that node $C$ has the highest extended remaining
resources in Fig.~\ref{fig:ex_input(b)}.

% where $L(n^V)$ and $L(n^S)$ are the set of connected links to $n^{V}$ and
% $n^{S}$ respectively,  $\text{CPU}^{R}(n^{S})$ is the amount of residual CPU resources of
% $n^S$ and $\text{BW}^{R}(l^{S})$ is the amount of residual bandwidth resources of
%  $l^S$, which is $\text{CAP}^{S}(l^{S})\cdot(1-\lambda_{l^{S}})$.

% Then, we sort other VN nodes in the ascending order of distance
% from the node with the highest resource concentration.   

% sort the VN nodes in the ascending order of distance
% to the  

% Let $\set{T} = N^V$ and call the virtual node with the highest
% $D(\cdot)$ the \emph{root node}. Sort $\set{T}$ in ascending order of
% distance from the root node. Here, among the virtual nodes with same
% distance, a node with higher $D(\cdot)$ is located in front of the nodes
% with less $D(\cdot)$.

% Sort the substrate nodes in the descending order of the $Q(\cdot)$
% value.  Choose $K$ substrate nodes with the highest $Q(\cdot)$ for some
% constant $K$. Denote each node by $n^{Q}_1, n^{Q}_2,\ldots,n^{Q}_K$. 

% {\bf Example using Figure~\ref{fig:ex_input}.}

\subsection{Searching Candidate Embeddings} \label{sec:search}

\noindent{\bf Overview.} 
In this subsection, we elaborate on {\bf Step 3}, where, starting with a given root SN
node in an embedding candidate, we finish the embedding candidate by
mapping other remaining VN nodes and links. The overview of the process
is as follows: We define a notion of {\em influence weight} assigned to all SN
links (to quantify interference in their neighborhoods). We then map 
the VN nodes to the SN nodes, so that the required link bandwidth between two VN nodes are
satisfied as well as minimize the aggregate influence weight of the mapped
{\em SN path}. This process is a joint link and node embedding that takes into account
our intention to prefer less interfering regions with higher capacity
(and thus more efficiently utilizing SN resources).
 
%  Once the two end nodes of a VN link is
% embedded, those of the SN path that will serve the VN link are also
% determined.  We choose the SN node that will serve a VN node by
% considering the link embedding stage of VN links connected to the VN
% node.

\smallskip
\noindent{\bf Influence weight and distance.}
The \emph{influence weight} $d_I(l)$ for an SN link $l$ is the
number of its interfering links (including itself) divided by
$\text{CAP}^S(l)$, \ie, $d_I(l) = (d_{l} + 1)/\text{CAP}^{S}(l).$ Then,
the influence weight for a {\em path} is the sum of influence weights of all
links in the path. The {\em influence distance} of two SN nodes is the minimum
influence weight among the paths connecting those two SN nodes. 
Fig. \ref{fig:ex_input}(c) illustrates an example of the influence weights. 
% In our algorithm, we sequentially embed VN nodes, where for a newly
% added VN node $v$, all the link band  

% We also define the influence length of
% a path by the sum of influence lengths of all links consisting of the
% path and the influence distance between two nodes is the length of the
% influence `shortest' path connecting two nodes.  Then, it would be
% reasonable to embed each VN link to corresponding influence shortest
% path.

\smallskip
\noindent{\bf Joint VN node and link mapping.} 
We now explain how the VN nodes are embedded in conjunction with the VN
links. Assume that we handle $i$-th VN node $n^V_i$ in the sequence by
{\bf Step 1}.  Let $\set{A}(n)$ be the set of already embedded VN nodes
adjacent to $n$ in $G^V.$ Assuming that the SN node for $n^V_i$ is
decided (whose rule will be explained shortly), the embedding of added VN links
between $n^V_i$ to all the VN nodes in $\set{A}(n)$ is done by the shortest
paths in terms of influence weight subject to each path has enough path
bandwidth to satisfy the bandwidth requirement of the corresponding 
VN link added due to node embedding.  

We now describe the rule for selecting the SN node $n^S_i$ which 
will embed $n^V_i$:
\medskip
\begin{separation}
  Embedding of $i$-th VN node $n^V_i$ with added VN links. 
\vspace{-0.4cm}
\separator
\vspace{-0.1cm}
% \noindent{\em - Selecting the SN node to embed}
\begin{eqnarray}\label{eq:increment}
&&  n^S_i  \in \arg\min_{n \in N^S}  \sum_{u \in \set{A}(n^V_i)}
\Big ( \text{BW}^{V}(\overline{u,n^V_i}) 
\times d_I(E(u),n) \Big ), \cr
&&\text{s.t. \ ``bandwidth requirements by added VN links,"}   
\end{eqnarray}
\smallskip
where $\overline{a,b}$ is the virtual
link between $a$ and $b,$ $E(u)$ is the SN node that serves
a VN node $u$ and $d_I(E(u),n)$ is the influence distance between $E(u)$
and $n$. 
% \smallskip
% \noindent{\em - Embedding the added links}
\end{separation}
\bigskip

The node embedding rule is designed to minimize the aggregate ``load stress''
added by embedding a new VN node. The load stress is measured by the
aggregate bandwidth requirement weighted by the influence distance over
the new SN paths. Weighting with the
influence weight is due to our design rationale of avoiding the region
with less capacity and high interference. 

% We introduce some notations for clear explanation.  $E(n^V)$ is the SN
% node to which a VN node $n^V$ is embedded.  $\set{H}(n^V)$ is the set of
% already embedded nodes among the neighbor of $n^V$.  $E(\set{H}(n^V))$
% represents the set of SN nodes to which VN nodes in $\set{H}(n^V)$ are
% embedded.  Let $\set{J}(n^V)$ be the set of VN links connecting $n^V$
% and a node in $\set{H}(n^V)$.

% We choose the SN $n^{\star}$ for $n^V$ which has enough CPU
% resources and correspond to the following :
% \begin{equation}\label{eq:increment}
%    \text{arg} \min_{n^S}  \sum_{u^V \in \set{H}(n^V)} \text{BW}^{V}(l(u^V,n^V)) \cdot d_I(E(u^V),n^S) 
% \end{equation}
% where, $d_I(E(u^V),n^S)$ is the influence distance between $E(u^V)$ and
% $n^S$ and $l(u^V,n^V)$ is the virtual link connecting $u^V$ and
% $n^V$. 
% Note that a bandwidth demand is also considered for more precise
% reflecing the effect of each VN link embedding in
% (\ref{eq:increment}). In order words, the $n^{star}$ has the minimum
% weighted sum of influence distances from SN nodes that $n^V$'s neighbor
% are previously embedded to.

\begin{figure}[]
  \centering 
  \subfigure[ ]{ % 
  \label{fig:ex_process(a)}
  \includegraphics*[width=0.3\columnwidth]{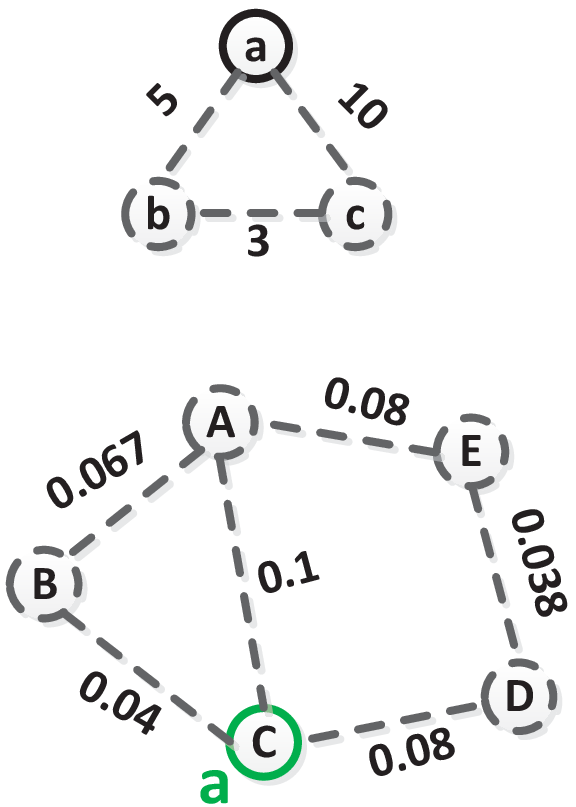}}
\subfigure[ ]{%
  \label{fig:ex_process(b)}
  \includegraphics*[width=0.3\columnwidth]{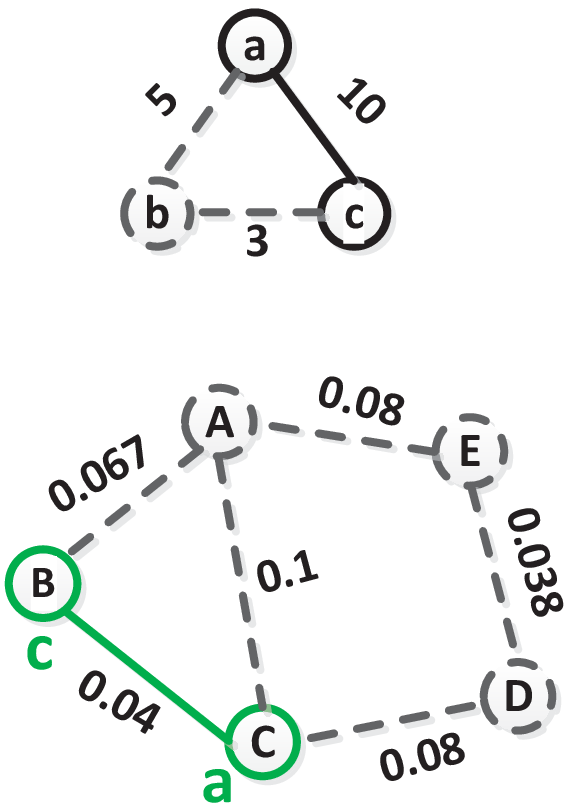}}
  \subfigure[ ]{%
 \label{fig:ex_process(c)}
\includegraphics*[width=0.3\columnwidth]{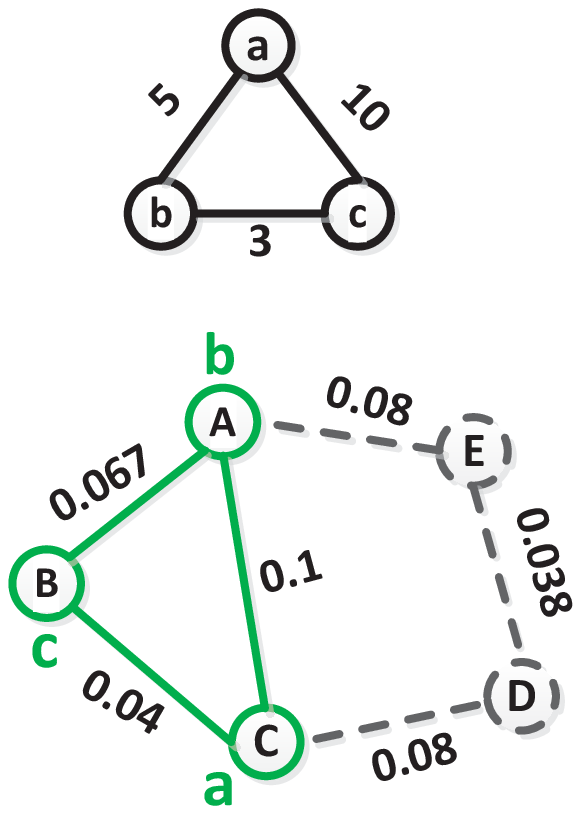}}
 \caption{Example of making a candidate embedding}
\label{fig:ex_process}
\vspace{-2mm}
\end{figure}

\smallskip
\noindent{\em Example.} 
We now illustrate the process using Fig.~\ref{fig:ex_process} that
magnifies the algorithm of this subsection from the example in Fig~\ref{fig:ex_input}. 
We assume that we consider a candidate embedding, where the first VN
node $a$ and the root SN node is $C.$ 
Since the next VN node is $c,$ we find the SN node which has the
shortest influence distance from $C$, which is node $B$. $c$ is embedded
to $B$ and VN link $(a,c)$ is mapped to link $(C,B)$
(Fig.\ref{fig:ex_process}(b)). We next search the SN node which will
serve node $b$. We calculate the (bandwidth requirement) weighted sum of
influence distances from node $C$ and $B$ to each of nodes $A$, $D$, and
$E$. That is: for $A$, $5\cdot 0.1 + 3\cdot0.067 = 0.701$, for $D,$
$5\cdot 0.08 + 3\cdot (0.04 +0.08) = 0.76$ and finally for $E,$ $5\cdot
(0.08 + 0.038) +3\cdot (0.067 + 0.08) = 1.031$. Thus, node $b$ is
embedded to node $A,$ and link $(a,b)$ and $(c,b)$ are embedded to link
$(C,A)$ and $(B,A)$, respectively (Fig. \ref{fig:ex_process}(c)).

\subsection{Comparison Metric} \label{sec:metric}

$K$ candidate embeddings are now ready, and we are in the stage of
quantifying  their qualities based on a metric which we explain in this
subsection ({\bf Step 4}). For a candidate embedding $E,$ we define the
embedding comparison metric $\sigma(E)$:
\begin{eqnarray*}
  \sigma(E) \triangleq \sum_{l \in L^S} (d_l+1) \times \lambda_l(E),
\end{eqnarray*}
where $\lambda_l(E)$ is the potential normalized load of link $l$ for $E,$ and
recall that $d_l$ is the number of interfering links with $l.$
Note that link $l$ forbids $d_l$ links to be activated at least
$\lambda_l$ portion of the time. The potential normalized loads weighted by $d_l+1$
captures the amount of offered loads considering interference (including
itself). Following such intuition, we prefer a candidate embedding with smaller value of $\sigma.$

% More details on the multiplication between $\lambda_l$ and $d_l$ is
% given here. 
% Several comparison metrics that evaluates embeddings' qualities
% in wireless networks can be proposed. One may consider the following principle of
% comparison among multiple embeddings: Embeddings that lead to less
% aggregate offered loads and interferences are preferred. From above intuitive principle, we propose the following comparison metric:
% $|\bm{\lambda}^T \bm{D}|,$ where $\bm{\lambda} = (\lambda_l : l \in
% L^S)$ and $ \bm{D} = (d_l: l \in L^S)$.
% Recall that the normalized loads $\lambda_l$ count also the required
% bandwidth of a candidate embedding as well as that of being served VN requests.
% Thus, it differs according to a candidate embedding.
% Link $l$ forbids $d_l$ links to be activated at least $\lambda_l$
% portion of the time. Thus, we have $\sigma(E) = \sum_{l}(\text{Req}(l)/\text{CAP}^S(l))\cdot
% d_l,$ which reflects the aggregate potential normalized loads.  For
% example, suppose that two candidate embedding $E1$ and $E2$ lead
% nomalized load $\bm{\lambda_1}$ and $\bm{\lambda_2}$ respectively. We
% decide that $E1$ is better than $E2$, if $\|\bm{\lambda_1}^T \bm{D}\| <
% \|\bm{\lambda_2}^T \bm{D}\|$.

The metric $\sigma$ is devised to be {\em fully compatible} with our joint node and
link embedding algorithm in Section~\ref{sec:search}.  Suppose that a VN
link $l^{V}$ is embedded to an SN path $\mathcal{P}$.  Then, the
increment of $\sigma$ is given by: 
\begin{eqnarray*}
\sum_{l \in \mathcal{P}} \frac{\text{BW}^{V}(l^V)}{\text{CAP}^S(l)} \cdot (d_{l}+1) 
&=& \text{BW}^{V}(l^V) \sum_{l \in \mathcal{P}}
\frac{d_{l}+1}{\text{CAP}^S(l)} 
= \text{BW}^{V}(l^V) \sum_{l \in \mathcal{P}} d_I(l).
\end{eqnarray*}

Comparing (\ref{eq:increment}) and the equation above, we observe the
equivalence in this equation. In other words, the searching process in
Section~\ref{sec:search} tries different regions to search $K$
embeddings by starting from the root nodes in different regions, and
their quality comparison is based on the metric that is instilled in
node and link embedding. 

\begin{figure}[]
  \centering
  \includegraphics*[width=0.9\columnwidth]{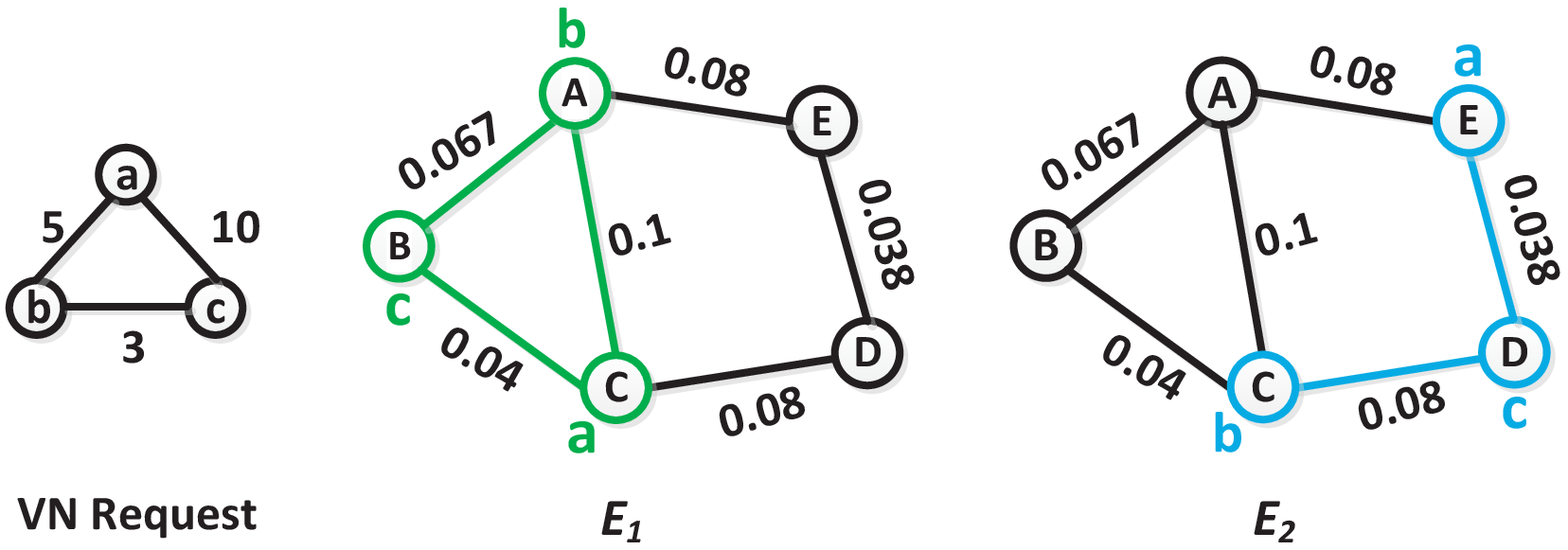}
  \caption{Two candidate embeddings for a given VN request}
  \label{fig:compex}
\end{figure}

As an example, consider two candidate embeddings $E_1$ and $E_2$ in
Fig.~\ref{fig:compex}. Note that virtual link $(a,b)$ is embedded to
path $(E,D,C)$ in $E_2$. Since $\sigma(\cdot)$ can also be expressed
by $\sum_{l \in L^S} Req(l) \times d_I(l)$,  $\sigma(E_1) = 10\cdot 0.04
+ 5\cdot 0.1 + 3\cdot 0.067\ = 1.101$ and $\sigma(E_2) = (10+5)
\cdot 0.038 + (3+5) \cdot 0.08 = \ 1.21.$ Thus, $E_1$ is preferred by the
metric.

% if VN node $n^V$ is embedded to $N^S$, the summation term in
% (\ref{eq:increment}) is the minimum increment of the metric value we can obtain 
% with respect to embedding of virtual links in  $\set{J}(n^V)$. In this
% sense, the chosen SN node $n^{\star}$ in (\ref{eq:increment}) is optimal for $E(n^V)$ under the
% given $E(\set{H}(n^V))$.

\subsection{Checking Feasibility}
\label{sec:FCmethod}
We finally select the embedding with the highest quality metric, which
passes feasibility check ({\bf Step 4}).  We now provide two candidate
ways of checking feasibility.

\smallskip
\noindent{\bf Exploiting Sufficient Conditions.} 
As discussed in Section~\ref{sec:challenge}, checking feasibility is
computationally intractable. One can apply a sufficient condition for a
given potential normalized load (recall its definition in
Section~\ref{sec:feasibility_checking}).  A well-known sufficient
condition is that for any link $l$ (a vertex in the conflict graph), the
sum of normalized loads of $l$ and $l$'s connected vertices in the conflict
graph is less than or equal to 1 (see \eg, \cite{gup07} for the formal
proof). Clearly, the fact that the sufficient condition is not met does
not imply that the tested embedding is infeasible. However, using this
sufficient condition is not a bad idea, because adding new requests
should be conservative such that existing virtualization service should
not be interrupted and also the approach of sufficient condition is
computationally attractive.

% If a candidate embedding meets the sufficient condition, the
% feasibility of it is guaranteed. On the contrary, if it does not we
% regard it as an infeasible embedding even though it is thing in reality.
% This sufficient condition can be applied any conflict graph, yet the gap
% from the optimal solution can be arbitrarily large.

\smallskip
\noindent{\bf Simulation via Smart Embedding.}
In spite of computational merit of the sufficient condition, its quality
can be bad for some network topologies, \ie, missing feasible
embeddings. Also, for a VN request, an embedding algorithm may not need
to produce the result very fast. If we spend a reasonable amount of
time, yet achieving more accurate feasibility check, the SN provider is
expected to earn larger revenue.

% \smallskip
% \noindent{\underline{\em Idea.}} 
We can examine embedding feasibility by actually simulating a MAC (or
its variant) for the potential normalized load $\bm{\lambda =
  (\lambda_l)}$.  In other words, we generate stochastic arrivals over
each SN link $l$, where the arrival mean is same as $\lambda_l.$
However, just performing simulation does not solve computational
intractability.  When $\epsilon=1,$ simulating the so-called {\em
  Max-Weight} can provides us with the result of feasibility
check. However, it is widely-known that Max-Weight requires to solve an
NP-hard problem (MWIS: Maximum Weight Independent Set problem) at each
time instance.  Note that our underlying MAC is
$\epsilon$-throughput-optimal.  For a general $\epsilon >0,$ which
requires an $\epsilon$-approximate algorithm of MWIS, the technical
challenge is that MWIS does not allow PTAS (Polynomial Time
Approximation Scheme) \cite{erl01}.

To achieve efficiency in conjunction with a reasonable complexity,
(\eg, polynomial), we perform {\em smart embedding.} The idea of smart
embedding is to restrict the use of SN nodes and links so that the
conflict graph of the embedded substrate network satisfies a special
geometrical property --- {\em polynomially bounded growth}\footnote{ Let
  $G=(V,E)$ be a graph and $d(u,v)$ be the hop-distance between node u
  and v. Then $r$-neighborhood of a node $v$ is denoted by $\Phi(v,r) =
  \{ v \in V | d(u,v) \leq r \}$.  We say that $G(V,E)$ is a
  polynomially bounded growing (PBG) graph with a polynomial function
  $p(r)$, if $|\Phi(v,r)| \leq p(r)$ for any $v$ and $r$.}. Recall that
the embedded substrate network is a subgraph of $G^S$ consisting only of
nodes and links serving some VN requests.  With  polynomial growth, we can
find a polynomial time algorithm which arbitrarily approximates the
original problem. For example, we allow a suboptimality gap $\epsilon
>0,$ then the complexity, which is a function of $\epsilon,$ is
polynomial with network size. We refer the readers to, \eg,
\cite{erl01,nie06} for $\epsilon$-approximation algorithms of MWIS in
PBG graphs.% Note that many classes

\smallskip
\noindent{\bf Comparison.} 
Two methods have different design rationales: (i) sufficient
condition---using the entire space of SN and coarse feasibility checking
or (ii) simulation via smart embedding---using a limited space of SN but
finer feasibility checking at cost of increasing complexity.  It is
interesting how two methods perform, which will be presented in
Section~\ref{sec:evaluation}.

\section{Performance Evaluation} \label{sec:evaluation}

% \begin{figure*}[t!]
%   \centering
%   \subfigure[Different feasibility checking]{%
%   \label{fig:feasibility}
%   \includegraphics[width=0.23\textwidth]{feasibility}}
%  \subfigure[Different SN density]{%
%   \label{fig:density}
%   \includegraphics[width=0.33\textwidth]{density}}
%  \subfigure[Specific VN topologies]{%
%   \label{fig:treestar}
%   \includegraphics[width=0.33\textwidth]{treestar}}
%   \caption{Average revenue for 6 algorithms}
%   \label{fig:6algorithm}
% \end{figure*}

We develop a wireless embedding
simulator (available in public \cite{WEM}) to evaluate the proposed
algorithm and analyze the impact of its key features.

\subsection{Simulation Environment}

\begin{figure}[]
  \centering
  \subfigure[High]{%
  \label{fig:SN-H}
  \includegraphics*[width=0.23\columnwidth]{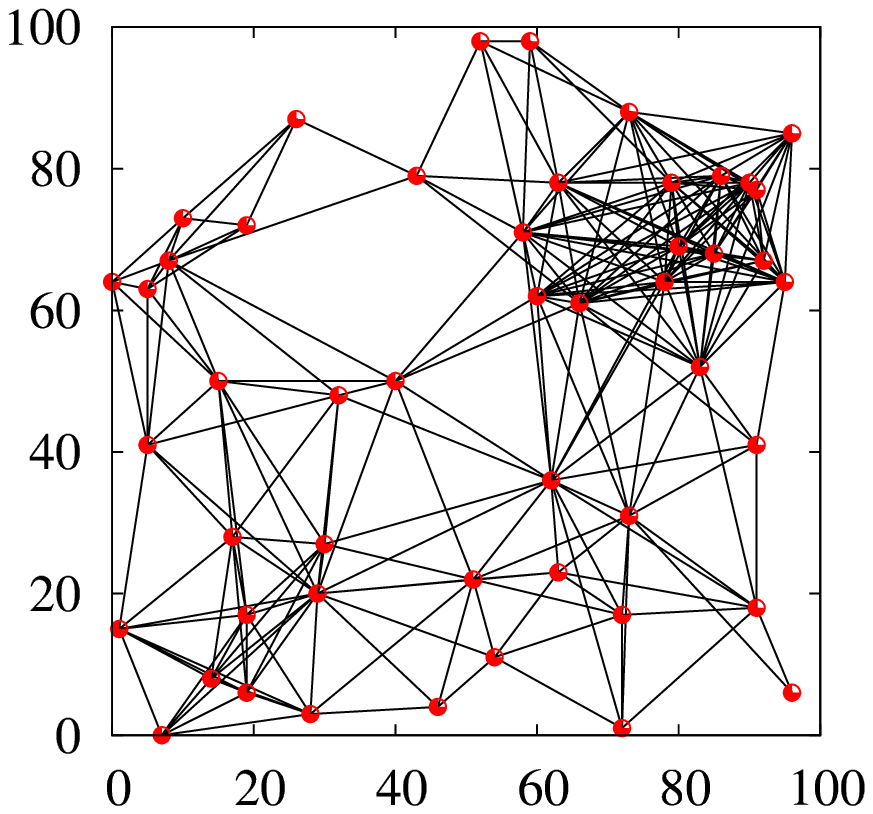}}%\hfill
  \subfigure[Middle]{%
  \label{fig:SN-M}
  \includegraphics*[width=0.23\columnwidth]{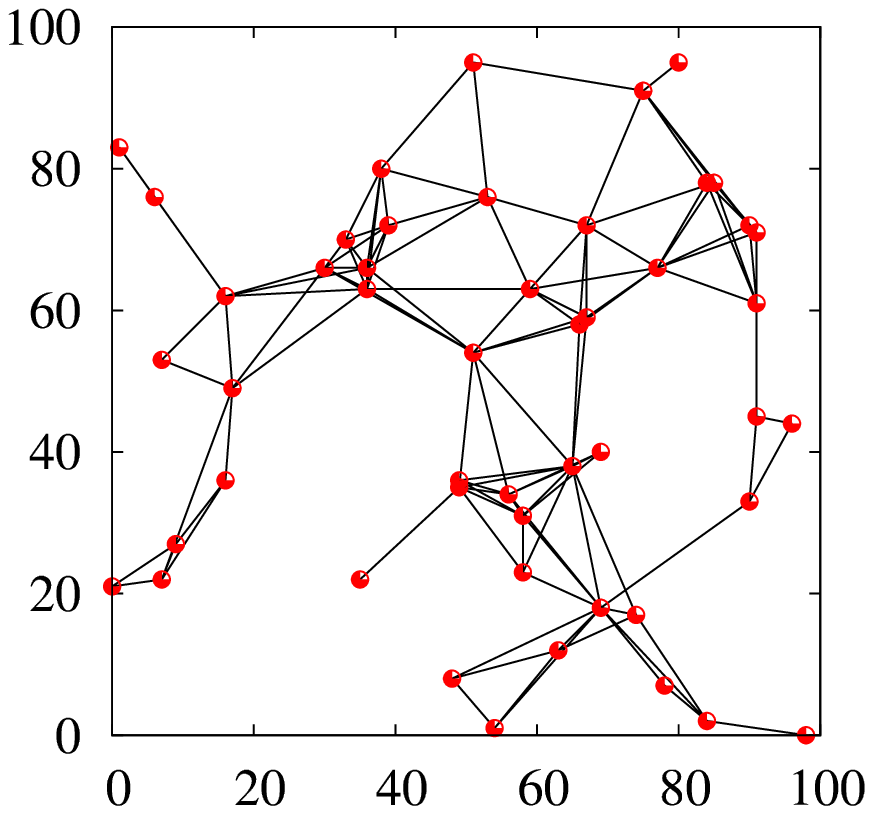}}
 \subfigure[Low]{%
  \label{fig:SN-L}
  \includegraphics*[width=0.23\columnwidth]{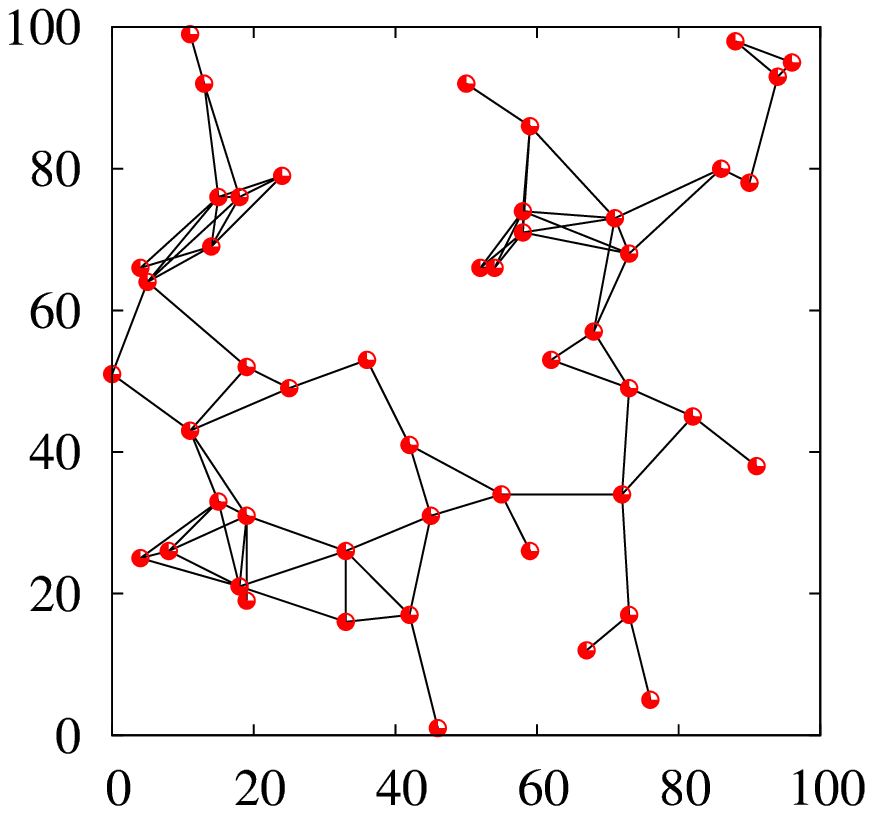}}%\hfill
  \subfigure[Grid]{%
  \label{fig:grid}
  \includegraphics*[width=0.23\columnwidth]{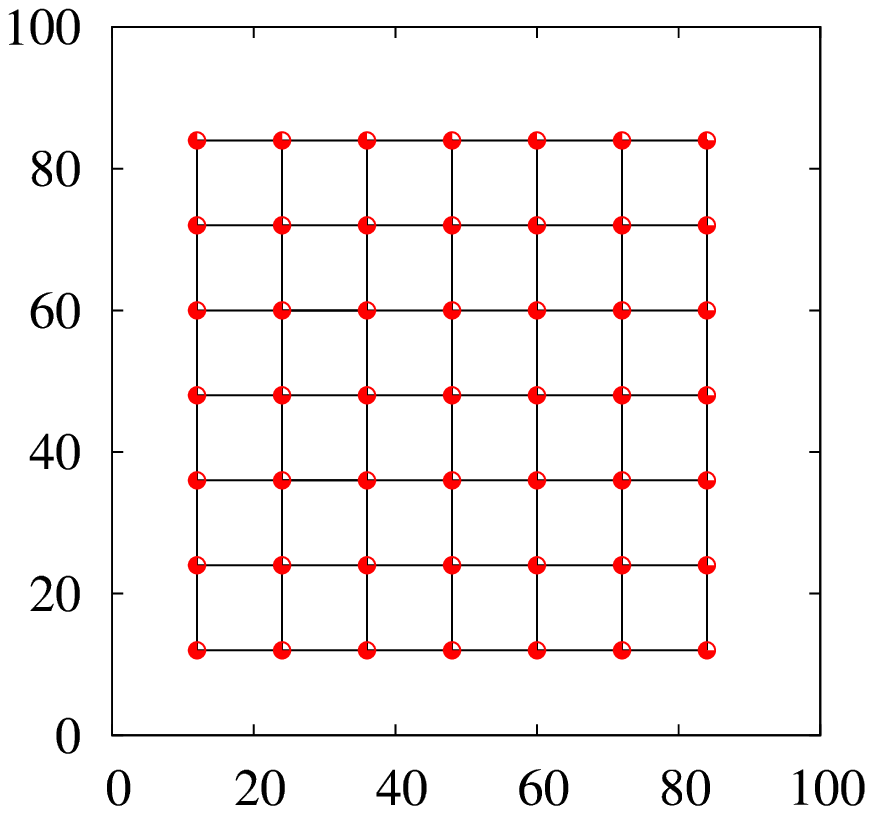}}
\caption{Tested substrate network topologies: (a), (b), and (c) are
  random instances with different densities. The grid topology includes
  49 (7$\times$7) SN nodes. In (a), (b), and (c), the transmission
  ranges are uniformly random over $[20,40],$ $[15,30],$ and $[10,20],$
  respectively.}
\label{fig:SNtopology}
\end{figure}

\noindent{\bf Substrate network.}  We set up a
grid square ($100 \times 100$), where an SN will be configured.  We
randomly place 50 substrate nodes on the grid square, where each node has
some random transmission range and any two nodes are connected if they
are within the transmission range of each other. In addition to this random
topology, we will show the results for other topologies, shown in
Fig.~\ref{fig:SNtopology}. We use random topologies with {\em middle}
density, unless explicitly specified.  To see the impact of SN
densities, we also vary the transmission range so that the number of SN
links ranges about 80 $\sim$ 240.
%  for all simulations,
% but we vary the SN densities to see the impact of SN density.  
% We connect a pair of nodes based on
% location and transmitting distance \cite{gup07,nie06}.
%One way of modeling a wireless network topology is to let wireless links be configured depending on the \emph{coverage area} of each node \cite{gup07,nie06}.
% For exmaple, the link between nodes $u$ and $v$ exists if and only if $u$ and $v$ can transmit to each other.
%Two nodes $u$ and $v$ are connected if $u$ is within the $v'$s coverage area and vice versa.
%We adjust the range of possible coverage radius so that we generate SNs that have various densities (see Fig.~\ref{fig:SNtopology}).
% We adjust the range of possible transmitting distance so that there are about 120 links on the SN.
The CPU resources of nodes and link capacities are set to follow a
uniform distribution between 100 and 300 units. In the revenue earned by
a VN request, we give higher priority to the links, where we choose
$\alpha =10.$ This choice is due to the fact that in wireless multi-hop
networks, link capacity is the more scarce resource, which has a larger
impact on the SN's revenue. A two-hop interference model, known to
suitably capture the MAC with RTS/CTS-like control messages, is adopted
in our evaluations. However, we verified through simulation (not
presented due to space limitation) that the overall trends do not
severely depend on interference models. We assume that $\epsilon =0.3,$
\ie, 30\% of MAC's suboptimality and overheads, which is used in
simulation-based feasibility check.

\smallskip 
\noindent{\bf VN requests.} In a VN request, we randomly select the number of VN nodes between 4 and 10. 
The probability of connecting a pair of virtual nodes is uniformly distributed over the interval
$[0.2,0.6].$ Each CPU resource requirement and bandwidth resource
requirement are also uniformly distributed between 1 and 10 units. The
arrival process of VN requests is modeled to follow a Poisson process
with an average of 5 per time window, unless explicitly specified. Each
VN request stays at the SN during the holding time following an
exponential distribution with a mean of 4 time windows.

\begin{figure}[]
  \centering
  \includegraphics*[width=0.33\textwidth]{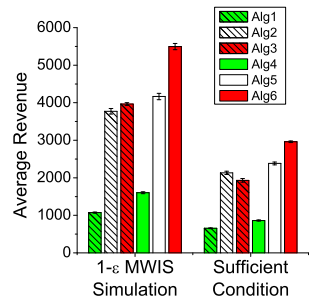} 
  \caption{Avg. revenue for feasibility check methods}
  \label{fig:feasibility}
\end{figure}

\begin{figure}[]
  \centering                                                                  
  \tabcolsep 0in
  \begin{tabular}{ccc}
    \includegraphics*[width=0.45\textwidth]{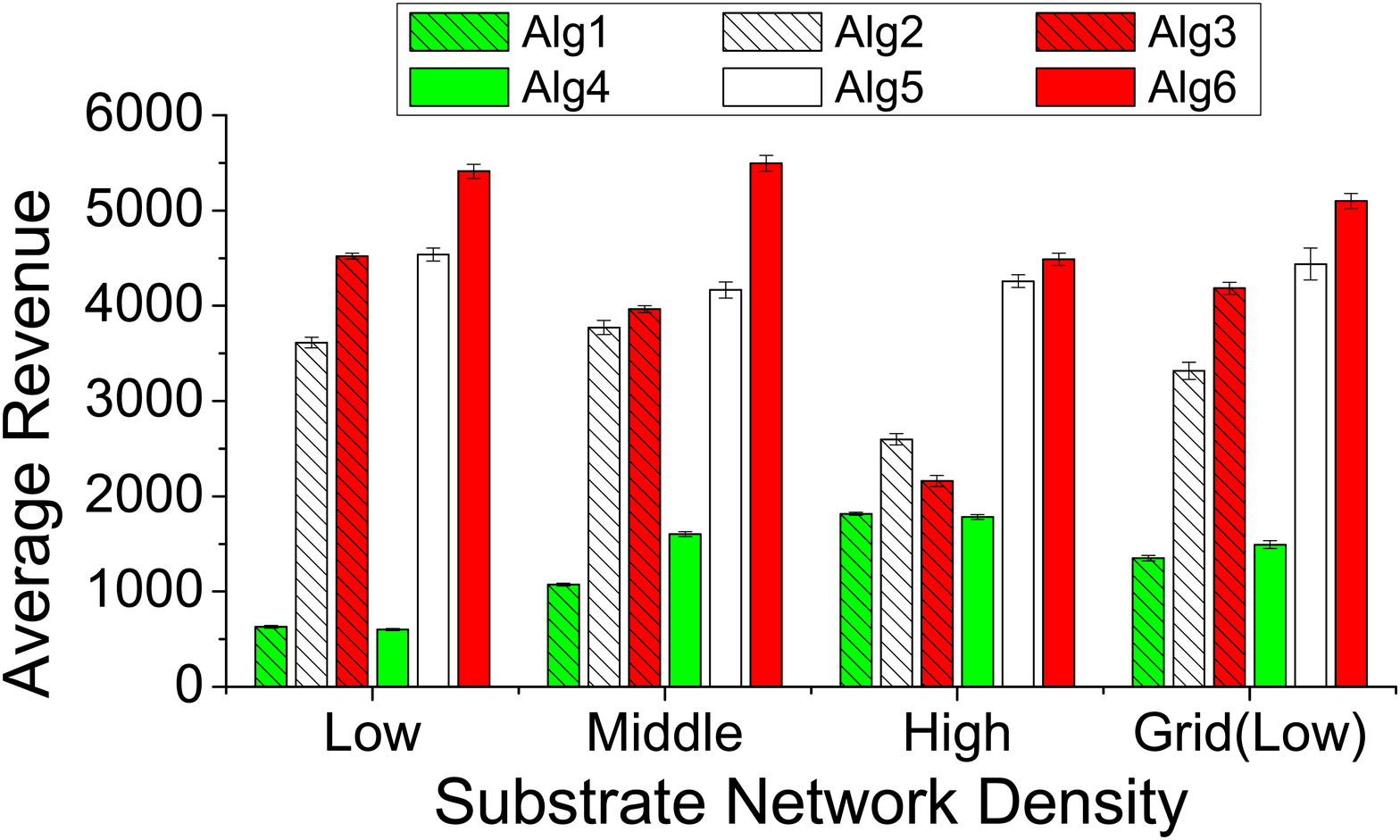}&
    \includegraphics*[width=0.38\textwidth]{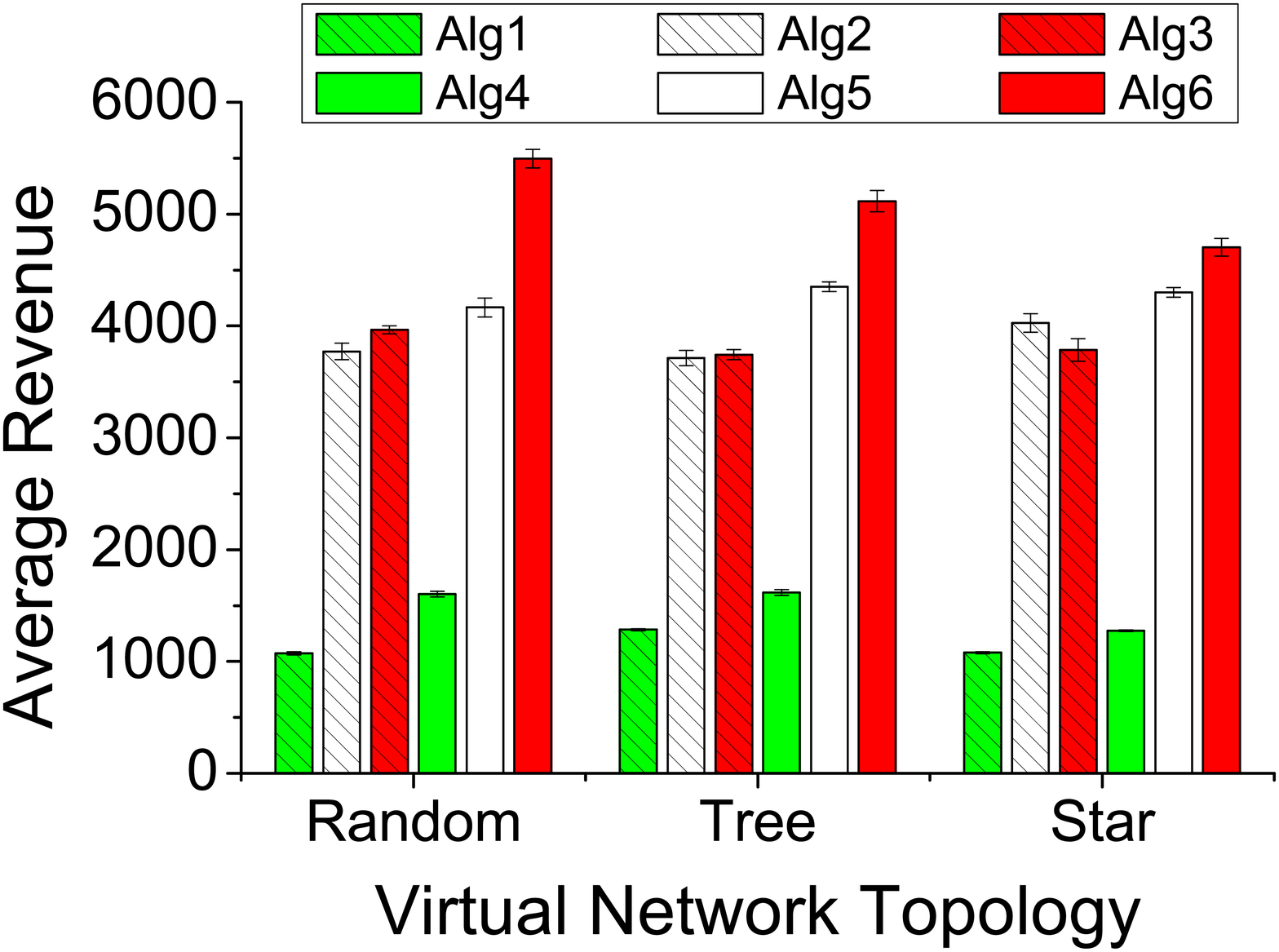}        &
    \includegraphics*[width=0.35\textwidth]{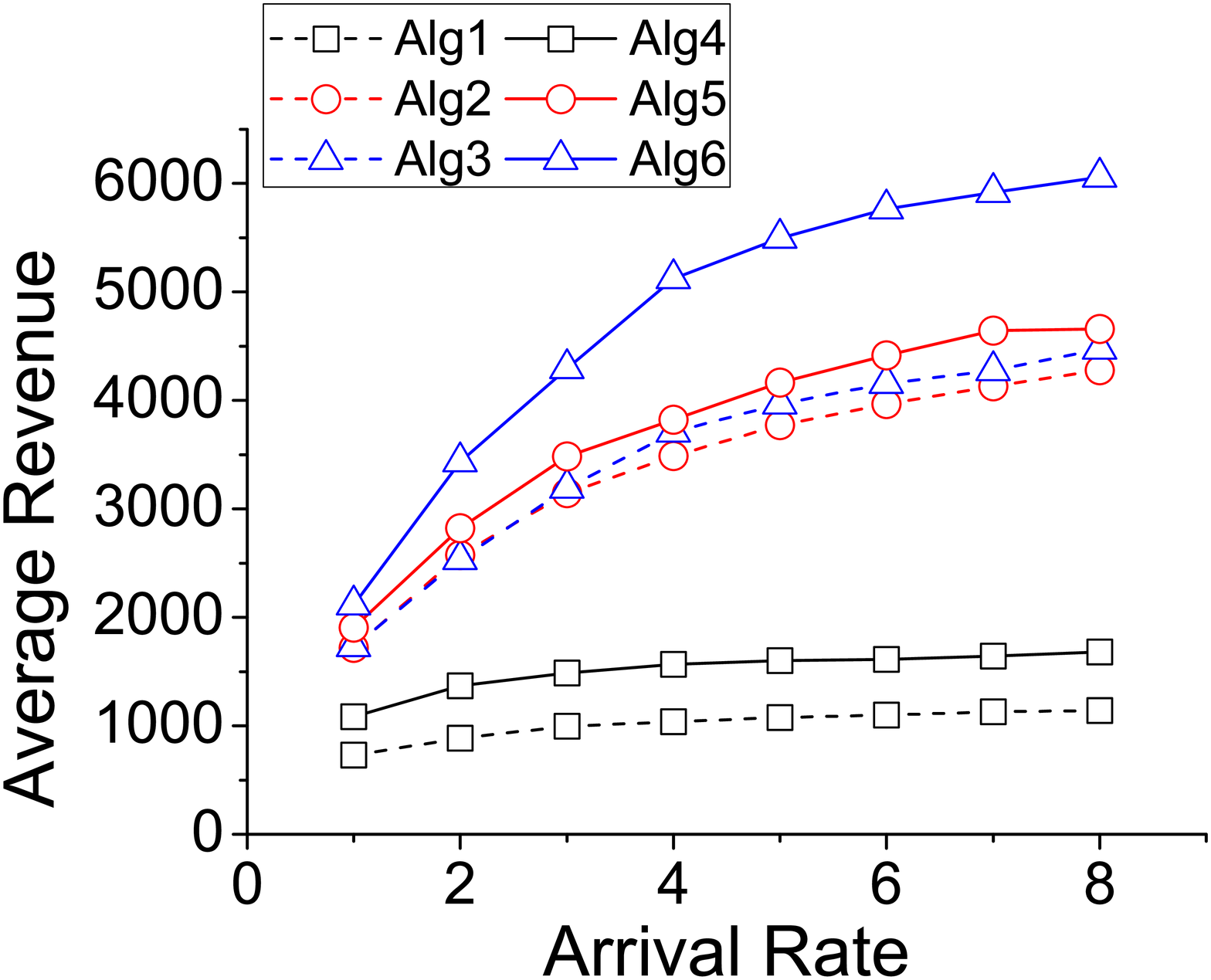} \\
    \parbox[t]{0.3\textwidth}{\scriptsize
      (a) Avg. revenue for SN densities}&
    \parbox[t]{0.29\textwidth}{\scriptsize
      (b) Avg. revenue for VN topologies}&
    \parbox[t]{0.25\textwidth}{\scriptsize
      (c) Avg. revenue for VN req. arrival intensities} \\
  \end{tabular}
  \caption{Comparison of tested algorithms for different scenarios}
  \label{fig:comparison}
\end{figure}

\smallskip
\noindent{\bf Tested algorithms.} To the best our knowledge, there does
not exist competitive embedding algorithms in wireless multi-hop
networks. However, we devise some possible algorithms to provide the
readers fair comparisons. In all algorithms, we select 8 candidate
embeddings for each VN request (\ie, $K=8$), The reason for this value
will shortly be provided in the simulation results.  We tested six
algorithms to see the effect of the key features in our algorithm, as
summarized in Table~\ref{tab:message}.

\begin{table}[]
\begin{center}
\caption{\label{tab:message} Tested algorithms: Alg6 corresponds to our
  proposed algorithm} 
  \begin{tabular}{|c||c|c|c|} \hline
    {\bf Link weight}&\multicolumn{3}{|c|}{\bf Link/node coupling} \\ \hline \hline
   & none& intermediate & full \\ \hline
  none& {\em Alg1} & {\em Alg2} & {\em Alg3}\\ \hline
   interference& {\em Alg4}&  {\em Alg5} & {\bf \em
     Alg6 (ours) }\\ \hline
\end{tabular}
\end{center}
\end{table}

Recall that there are two main features of our algorithms: (F1) joint
node/link embedding, and (F2) a notion of influence weight that
quantifies a link weight which is used to select an embedded SN path.
Regarding (F1), we artificially make three classes ({\em none}, {\em
  intermediate}, and {\em full}) based on existing algorithms in wired
networks, and two classes for (F2) ({\em none} and {\em
  interference}). In (F1), {\em none} corresponds to the greedy
algorithm in \cite{yu08}. {\em Intermediate} corresponds to a slightly
modified algorithm in \cite{zhu06}.  Both {\em intermediate} and {\em
  full}, share the feature that embedded SN nodes should be closely
placed. The difference is that in {\em intermediate}, the inter-distance
between VN nodes is not considered, whereas in {\em full}, two VN nodes
with shorter distance are embedded to two SN nodes with shorter
distance.  In (F2), no algorithms in wired networks consider
interference, \eg, \cite{lis09}. Thus, to purely focus on how the
influence weight affects the performance, we choose algorithms without
the influence weight as {\em none}, so the SN path is computed just by
considering the number of hops.

\subsection {Simulation Results} 

% We summarize the key messages of our results as follows.

% \smallskip
% \noindent{\bf \em Comparison for tested algorithms} 
% \smallskip

\noindent{\bf Impact of feasibility check methods.} 
We first consider the results of comparing two feasibility checking
methods, shown in Fig.~\ref{fig:feasibility}.  We observe that the
simulation-based method significantly outperforms the sufficient
condition based method (about two times). From this, we can see that
rather than fully utilizing SN resources with an inaccurate feasibility
check, it is more desirable to apply a strict checking method even with
a slightly limited usage of SN resource. We apply the simulation based
feasibility check to the rest of the simulations. Note that all tested
algorithms are equipped with the same feasibility checking method for
fair comparison.

\smallskip
\noindent{\bf Impact of SN density.}  
We now start to compare the performance of tested algorithms. We look at
the impact of different SN densities, shown in
Fig.~\ref{fig:comparison}(a).  In all graphs, {\em Alg6} outperforms
other algorithms. An interesting observation here is that {\em as SN
  density increases,} \eg, see the high density case, {\em the
  importance of considering interference in the link weight becomes
  stronger.} Thus, the performance gap between {\em Alg6} and {\em Alg5}
is small, whereas {\em Alg3}'s performance gap from {\em Alg6}
increases, compared to other lower SN density cases.

% We also simulate above algorithms over other SN topologies with
% different density (Fig.~\ref{fig:density}).  Through the simulation, we
% find that differences in revenue among algorithms become larger in
% denser SN.  This is mainly because a link of a dense network suffers
% heavy interference from other links.  Thus, the proposed algorithm that
% considers link interference can show its superiority more clearly in
% dense networks.
% OR
%Through the simulation, we observe that superiority of the coupling and influence weight is clearer in denser network.
%One challenge of embedding algorithm in wireless multihop networks is to embed links and nodes reducing interference.
%This means that a dense SN which has heavier interference is a proper environment to evaluate embedding algorithm.
%Therefore, the observation implies that the proposed algorithm is better to resolve interference than other algorithms.

\smallskip
\noindent{\bf Impact of VN topology.}  
VNs may often have special topological structures such as tree,
hub-and-spoke, and star. This topology depends on the type of
virtualization service. For example, a game service with a single server
is likely to form a star topology. We study this topological impact. 
Fig.~\ref{fig:comparison}(b) shows the avg. revenue comparison for
tree, star, and random topologies. We observe that in star topology,
four algorithms, {\em Alg2, 3, 5, 6}, do not lead to a large performance
gap. We analyze this observation as follows: In star topology, all links
are connected to a ``center'' VN node. Then, severe {\em local concentration}
is experienced around a node: all VN links should be embedded
to the paths concentrated around the SN node which embeds the center VN
node. This load concentration prevents a big star topology from being
embedded in all algorithms, and only small star topologies are accepted.
This trend is supported by the results for the tree
topologies, whose degree of concentration is between random and star
topology, where the performance gap is in between those two topologies.
However, we still observe that joint consideration of node and link is
crucial in improving the performance.

%  which may behave as the bottleneck links with scarce
% resource.    

% for being accepted.  Conversely,
% we can infer that VN link in accepted VNs is embedded to `link' rather
% than `path'.  As a result, the merit of using influence weight is
% diminished.

\smallskip
\noindent{\bf Impact of VN arrival intensity.}  
We also vary the VN arrival intensities by testing various arrival
rates, ranging from 1 to 8, as shown in
Fig~\ref{fig:comparison}(c). First, in all algorithms, the revenue
curves are concave with ``diminishing returns.'' This is because for low
arrival rates, bigger VN topologies (and thus, bigger revenues) can
dominate the total earned revenue. However, for high arrival rates, in
addition to those big VNs, only small VNs take effect in increasing
revenues. Second, {\em Algs1-6} are grouped into three classes in terms
of average revenue performance, where (i) joint link/node embedding is
crucial, (ii) interference-aware link weights leads to additional
revenue increase. This implies the importance of the two key features in
{\em Alg6}.

\begin{figure}[]
  \centering
  \subfigure[Avg. revenue]{%
  \label{fig:can-rev}
  \includegraphics*[width=0.5\columnwidth]{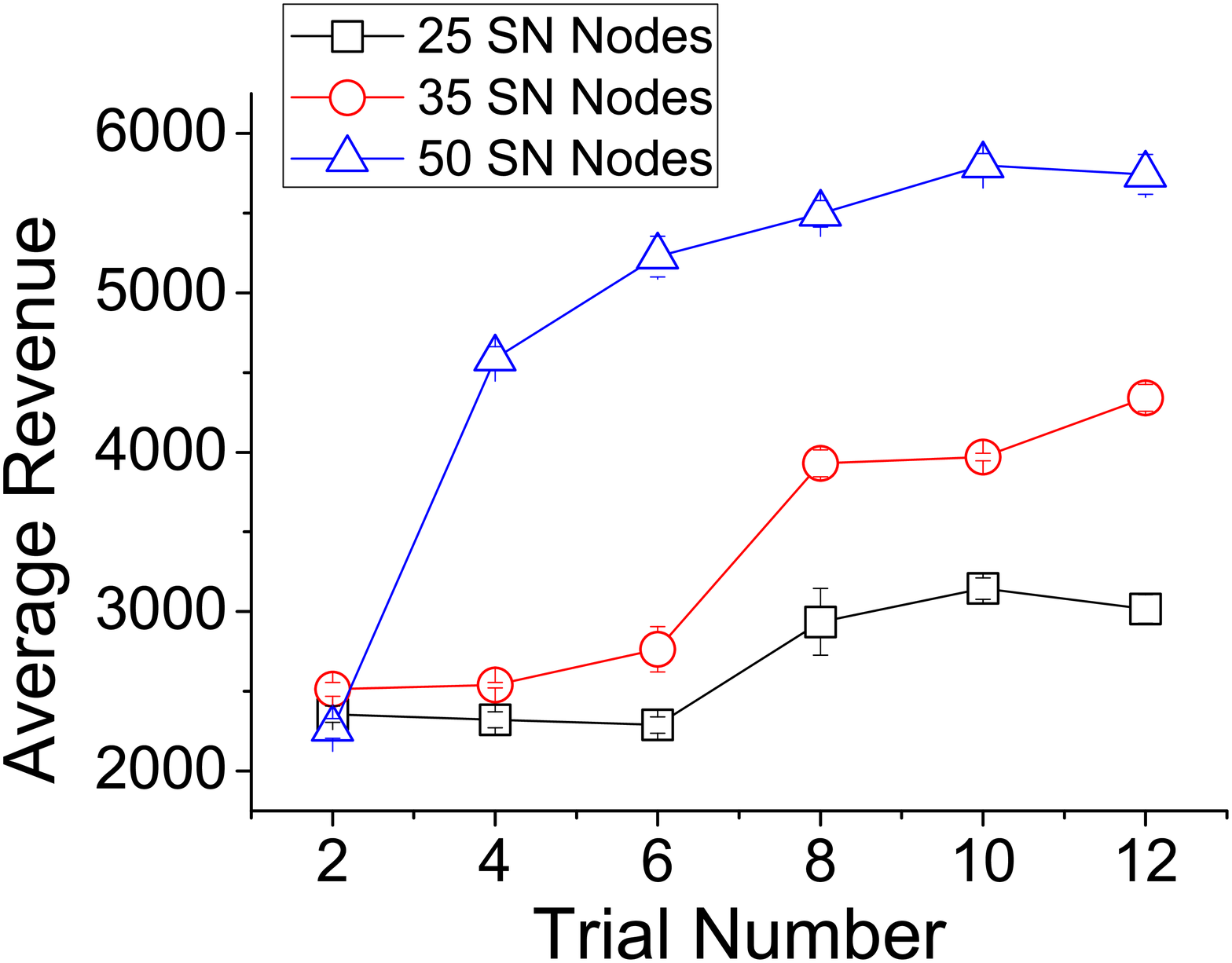}}
 \subfigure[Running time per VN request]{%
  \label{fig:can-time}
  \includegraphics*[width=0.43\columnwidth]{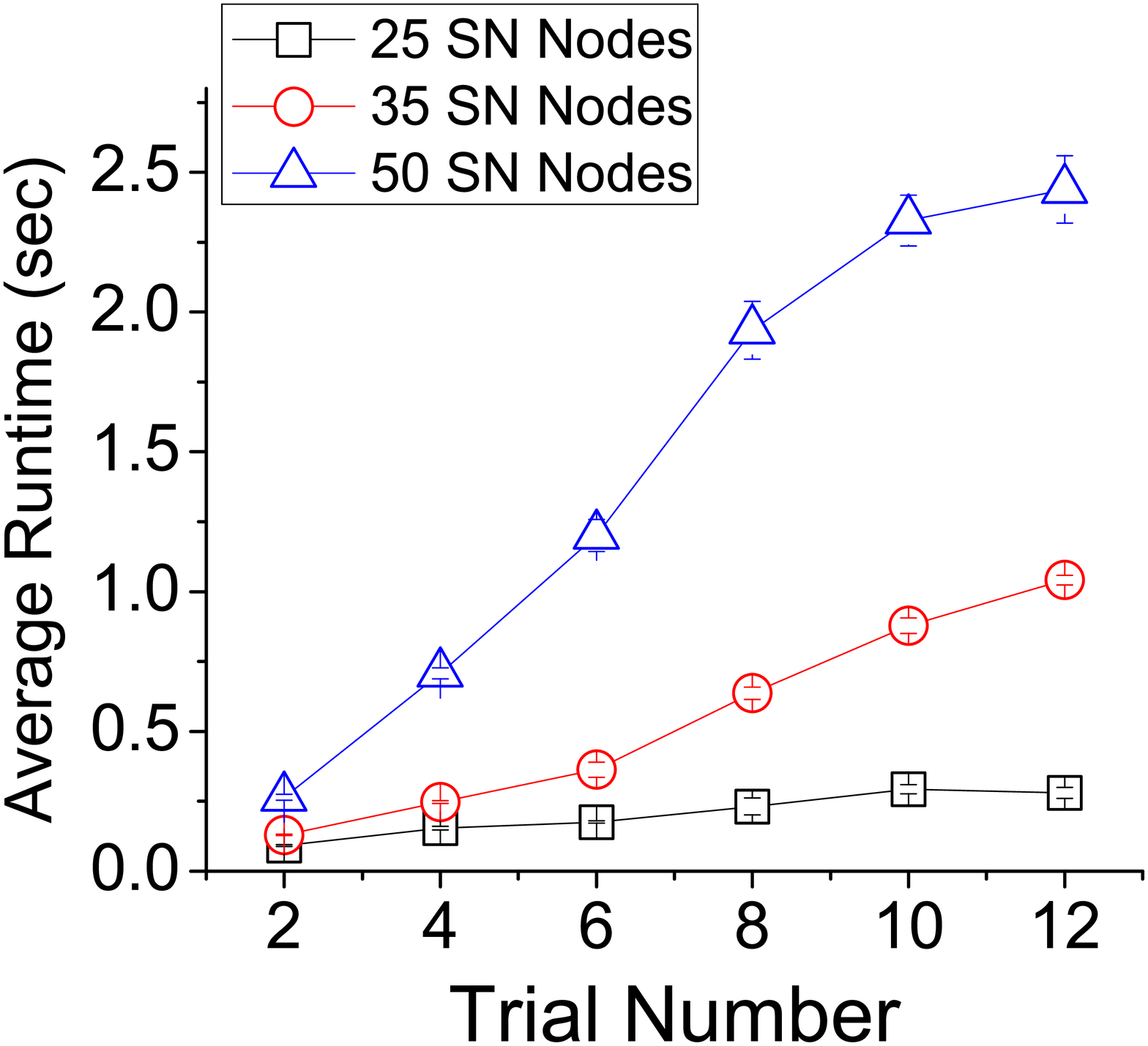}}%\hfill
\caption{Effect of the number of considering candidates}
  \label{fig:candidate}
\end{figure}

\smallskip
\noindent{\bf Impact of search number $K.$} 
The search number should strike a balance between the running time and
efficiency. Fig.~\ref{fig:candidate} shows the average revenue as well
as running time as $K$ increases. We vary the number of SN nodes
using {\em middle} SN density.
% The parameter $K$ essentially
% characterizes the number of regions in the SN.  The optimal search
% number depends on many things, but we note two major parameters, the
% number of SN and VN nodes. This is intuitive, because as the number of
% SN nodes increases, it is necessary to search more. Conversely, a larger
% number of VN nodes lets us search less. Thus, a heuristic rule of
% choosing $K = |N^S|/ |N^V|.$ For example, in a simulation of
% Fig.~\ref{fig:candidate} with $|N^S| = 50,$ $K = 50/7 \approx 7.1.$ In fact,
Fig.~\ref{fig:candidate} shows that after $K=8,$ the revenue saturates,
whereas the computation time increases linearly. The computation time also increases more
sharply with the increasing number of SN nodes, because the number of SN links also
increases due to the fixed SN density. The choice of this search number
$K$ may depend on the SN and VN sizes. The simulation results
imply that a small search number out of the huge search
space may be sufficient enough. Fig.~\ref{fig:candidate} also implies that our algorithm
is computationally tractable, because the algorithm generates the
embedding results in the order of seconds, which is reasonable
in practical applications.

\smallskip
\noindent{\bf Summary.} First, the effect of joint node/link embedding
is large. The node embedding in a ``non-coupling'' algorithm does not
consider the neighborhood interference of SN nodes, and thus, with a
limited search number, it is highly likely to choose inefficient
embedding candidates. Second, a more accurate feasibility check is
necessary for efficient embedding.  This accuracy comes with the cost of
additional time, but it is still reasonable in practical cases.  Third,
considering wireless interference as the link weight is also important,
where in some cases, the performance gap amounts to about 35\% for the
algorithms that are unaware of interference.  Fourth, VN topology may
significantly impact the effect of embedding algorithms, especially in
wireless multi-hop networks, due to resource concentration at nodes that
leads to the generation of bottlenecks. % Fourth,

\section{Concluding Remarks}
\label{sec:conclusion}

In this paper, we propose an embedding algorithm over wireless multi-hop
networks. The key challenges in wireless embedding originate from
inter-link interference, which makes the issues of feasibility check and
candidate embedding search. The main features of the proposed algorithm
are joint node/link embedding and interference-aware link weight.  Our
proposed algorithm may leave some rooms for further improvement, but our
findings on the key features are expected to provide useful implications
to the embedding algorithm research in wireless multi-hop networks. One
can extend our algorithm to more practical MACs, \eg, 802.11 DCF, where
our key ideas can be utilized except for feasibility check. Although it
is not the scope of this paper, feasibility check in 802.11 may be
borrowed from many research papers on admission control in 802.11-based
multi-hop networks, see \eg, \cite{HT09}, which is left as an
interesting future work.

{
\bibliographystyle{model1-num-names}
\bibliography{all}
}
\end{document}